\documentclass[12pt,prd,aps,epsfig,floats,preprint,eqsecnum,nofootinbib]{revtex4}

\usepackage{color}
\usepackage{amsmath}
\usepackage{amsfonts}
\usepackage{graphicx}
\usepackage{subfigure}
\usepackage{rotating}
\usepackage{etoolbox}
\usepackage{fontenc}
\usepackage{textcomp}
\usepackage{xspace}
\usepackage{siunitx}
\usepackage{graphicx,amsmath, nccmath, amssymb,epsfig,verbatim,mathrsfs,array,layout,latexsym}
\usepackage{multirow}
\usepackage{hyperref}
\renewcommand\thesubsection{\arabic{subsection}}

\usepackage{tikz-feynman}
\usepackage{tikz}
\usetikzlibrary{arrows}
\usepackage[compat=1.1.0]{tikz-feynhand}

\tikzfeynmanset{
  fermion/.style={
    /tikz/postaction={
      /tikz/decoration={
        markings,
        mark=at position 0.5 with {
          \arrow{>[length=6pt, width=5pt]};
        },
      },
      /tikz/decorate=true,
    },
  },
}

\begin{document}
\setlength{\parindent}{0pt}

\newcommand{\pythia}{\textsc{Pythia}~8.3\xspace}
\renewcommand{\mg}{{\sc\small MadGraph5\_aMC@NLO}}
\newcommand{\smeftsim}{\texttt{SMEFTsim}}
\newcommand{\delphes}{\textsc{Delphes3}\xspace}
\newcommand{\EFTfitter}{EFT\textit{fitter}\xspace}
\newcommand{\bat}{BAT.jl\xspace}

\newcommand{\sW}{\sin \theta_W}
\newcommand{\cW}{\cos \theta_W}

\DeclareSIUnit{\ab}{\text{ab}}

\newcommand{\emiss}{E_T^{\text{miss}}}

\title{Total Drell-Yan in the flavorful SMEFT}

\author{Gudrun~Hiller}
\email{ghiller@physik.uni-dortmund.de}
\affiliation{TU Dortmund University, Department of Physics, Otto-Hahn-Str.4, D-44221 Dortmund, Germany}
\author{Lara~Nollen}
\email{lara.nollen@tu-dortmund.de}
\affiliation{TU Dortmund University, Department of Physics, Otto-Hahn-Str.4, D-44221 Dortmund, Germany}
\author{Daniel~Wendler}
\email{daniel.wendler@tu-dortmund.de}
\affiliation{TU Dortmund University, Department of Physics, Otto-Hahn-Str.4, D-44221 Dortmund, Germany}

\begin{abstract}
We perform a global analysis of Drell-Yan production of charged leptons and dineutrinos, the latter in missing energy plus jet events, in proton-proton collisions within 
the Standard Model Effective Field Theory (SMEFT).
The combination allows for the removal of flat directions, sharper limits and to probe more couplings than the individual observables, which we
show performing a fit to LHC-data.
We also find that limits have only mild dependence on lepton flavor patterns; hierarchies in quark flavors are driven by the parton distribution functions.
The strongest constraints are on couplings involving the first and second generation quarks, exceeding 10 TeV.
Combining flavor and high-$p_T$ data, the limits on electroweak and gluon dipole operators can be improved, by up to a factor of three, highlighting once more that a more global approach increases sensitivities significantly.
We also estimate the improvements in reach over existing data for the high-luminosity LHC (HL-LHC) by $\sim 1.5$, and future collider options such as the high-energy LHC (HE-LHC) by $\sim 3$ and FCC-hh by $\sim 8$.
To maximize the new physics reach kinematic cuts and binning needs to be adjusted to each quark-flavor separately.
\end{abstract}

\maketitle

\tableofcontents

\section{Introduction}
\label{sec:introduction}

To study fundamental dynamics at the electroweak scale and beyond, the Standard Model (SM) of the strong and electroweak interactions is, given its successes~\cite{ParticleDataGroup:2024cfk},
a likely good benchmark theory to start. To learn when and where the SM eventually fails is informative and could facilitate progress towards deeper questions, such as the origin of dark matter, the matter-antimatter-asymmetry, or flavor. This endeavor requires directions from theory and model building, and a broad program for experimental searches.
A model-independent and systematic approach to further explorations is to treat new physics degrees of freedom from  above the weak scale effectively in towers of higher dimensional operators, encoded by the Standard Model Effective Theory (SMEFT)~\cite{Buchmuller:1985jz}.
As also proven for instance in Fermi's theory including applications to charm and $b$-physics~\cite{HeavyFlavorAveragingGroupHFLAV:2024ctg}, effective field theories are powerful tools as 
they allow for a global view, to correlate physics from different scales, sectors, observables and experiments.

In this work we consider dilepton production in $pp$-collisions, aka the Drell-Yan process, in the SMEFT. The $pp \to \ell^+ \ell^{(\prime) -}$ one into charged leptons $\ell, \ell^\prime=e, \mu, \tau$, which we term charged lepton Drell-Yan (CLDY),
has been previously studied, e.g. in Refs.~\cite{Dawson:2018dxp, Fuentes-Martin:2020lea, Boughezal:2022nof, Greljo:2021kvv, Greljo:2022jac, Allwicher:2022gkm, Boughezal:2023nhe,Allwicher:2024mzw,Crivellin:2021rbf,Corbett:2024evt}. Recently the production of a dineutrino pair and a tagging jet (MET+j), $pp \to \nu \bar{\nu} j$ has also been analyzed~\cite{Hiller:2024vtr}.
MET+j and 'classic' Drell Yan share many features: energy-enhancement of dipole and semileptonic four-fermion operators,
but also beneficial synergies are expected: while dipole operators involving the $Z$-boson are probed by both processes, only CLDY is sensitive to the photon dipoles,
and only MET+j accesses the gluon dipoles which are fully energy-enhanced.

We perform a global analysis of LHC-data on production of charged leptons and dineutrinos, and work out synergies also including flavor data.
We concentrate on flavor-changing neutral current (FCNC) couplings of the quarks to make contact with flavor physics. In this case interference terms between the SM and the SMEFT are suppressed since the SM is flavor-diagonal at leading order in quark mixing.

The plan of the paper is as follows:
In Section \ref{sec:SMEFT}, dilepton production, $\ell^+ \ell^{(\prime) -}$ and $\nu \bar \nu$, in proton-proton collisions in the SMEFT is reviewed.
We give flavor aspects and the sensitivity of cross sections to Wilson coefficients (WCs).
In Section \ref{sec:fit_data} we give the LHC-data used in the global fit, and describe the details of the fit procedure.
The outcome of the fit, including limits on WCs and synergies between charged lepton and dineutrino observables, are given in Sec \ref{sec:Results},
together with a comparison of results for lepton flavor universal, lepton flavor violating and democratic patterns.
In Section \ref{sec:Flavor} we work out the benefits of flavor constraints from rare hadron decays on the global analysis. 
Estimations for the High-Luminosity LHC (HL-LHC)~\cite{ZurbanoFernandez:2020cco}, a possible High-Energy LHC (HE-LHC)~\cite{FCC:2018bvk}, and the Future Circular Collider (FCC-hh)~\cite{Benedikt:2022kan} 100 TeV $pp$-collider are presented in Section \ref{sec:FutureCollider}.
We conclude in Section \ref{sec:conclusion}. Auxiliary information and figures are provided in the appendix.

\section{Dilepton production in the SMEFT}
\label{sec:SMEFT}

We briefly review the SMEFT formalism relevant for the charged and neutral leptoproduction in Sec.~\ref{sec:SMEFT_framework}
and give flavor aspects of our analysis. We begin with quarks in Sec.~\ref{sec:mass_basis}., followed by leptons in Sec.~\ref{sec:leptons}.
We discuss sensitivities to effective coefficients of the parton level cross sections together with the interplay and synergies of CLDY and the MET+j channels in Sec.~\ref{sec:cross_sections}.

\subsection{SMEFT framework and operators}
\label{sec:SMEFT_framework}

We focus on dimension-six SMEFT operators in the Warsaw basis~\cite{Grzadkowski:2010es}, with the Lagrangian~\cite{Buchmuller:1985jz}
\begin{equation}
  \mathcal{L} = \mathcal{L}_{\text{SM}} + \sum_i \frac{C_i}{\Lambda^2} O_i \,,
\end{equation}
where $\mathcal{L}_{\text{SM}}$ is the SM Lagrangian. The operators $O_i$ are composed out of SM fields, and respect Poincare and SM gauge symmetry,
$SU(3)_C \times SU(2)_L \times U(1)_Y$; the $C_i$ are the corresponding Wilson coefficients.
In SMEFT, the scale of new physics, $\Lambda$, has to be sufficiently above the electroweak scale, $v \simeq 246$ GeV.
We use $\Lambda = 1\,\text{TeV}$ for the numerical analysis.

Four-fermion operators can significantly impact dilepton production at the LHC, as their contributions to the cross sections are enhanced by the partonic center-of-mass energy as $\hat{s}^2/\Lambda^4$, where $\hat{s}$ is the square of the partonic center-of-mass energy~\cite{Farina:2016rws, Allwicher:2022gkm, Allwicher:2024mzw}. Furthermore, 
the MET+j process receives contributions from the gluon dipole operator subject to the same scaling with energy~\cite{Hiller:2024vtr}~\footnote{In appendix \ref{app:G_dipoles} we give further details on the energy enhancement of the gluon dipole operators. }.
Contributions from electroweak dipole operators involving quarks to dilepton cross sections are scaling as $v^2\,\hat{s}/\Lambda^4$. We therefore focus on the dipole and semileptonic four-fermion operators summarized in Tab.~\ref{tab:operators}.
While other classes of operators, such as $Z$-penguins and leptonic dipole operators contribute to dilepton production as well, they are already tightly constrained by LEP data~\cite{Efrati:2015eaa,Celada:2024mcf,Bellafronte:2023amz}. Note however, that they can be included in a global fit in a straight-forward manner.

We employ the following notation: $q_i$ and $l_\alpha$ are the $SU(2)_L$ doublet quark and lepton fields which, respectively, contain left-handed up- and down-type quarks, and the charged lepton and neutrino fields. In components, $q=(u^L,d^L)$ and $l=(\nu,\ell)$. The $u_j, d_j,$ ($e_\beta$) are the right-handed quark (lepton) singlets. The indices $ i, j $ label quark generations, while $\alpha, \beta$ are generation indices of the leptons, $i,j,\alpha,\beta=1,2,3$. The Higgs doublet is denoted by $\varphi$, while $\tilde{\varphi} = i \tau^2 \varphi^*$ denotes its charge-conjugate. The $\tau^I$ are the Pauli matrices, and $T^A = \frac{1}{2} \lambda^A$ are the generators of QCD.
The field-strength tensors $B_{\mu\nu}$, $W_{\mu\nu}^I$, and $G_{\mu\nu}^A$ correspond to the $U(1)_Y$, $SU(2)_L$, and $SU(3)_C$ gauge groups, respectively. The Levi-Civita symbol $\epsilon_{km}$ acts on $SU(2)_L$ indices $k,m$, with $\epsilon_{12} = +1$.

\begin{table}[h]
  \renewcommand{\arraystretch}{1.5}
  \setlength{\tabcolsep}{10pt}
  \centering
  \begin{tabular}{|l l | l l | l l|}
    \hline
    \multicolumn{6}{|c|}{{Dipole}} \\
    \hline
    $O_{\underset{ij}{uB}}$ & $\bigl(\bar q_i \sigma^{\mu\nu} u_j \bigr) \tilde{\varphi} B_{\mu\nu}$ & $O_{\underset{ij}{uW}}$ & $\bigl(\bar q_i \sigma^{\mu\nu} u_j \bigr) \tau^I \tilde{\varphi} W_{\mu\nu}^I$ & $O_{\underset{ij}{uG}}$ & $\bigl(\bar q_i \sigma^{\mu\nu} T^A u_j \bigr) \tilde{\varphi} G_{\mu\nu}^A$ \\
    $O_{\underset{ij}{dB}}$ & $\bigl(\bar q_i \sigma^{\mu\nu} d_j \bigr)\varphi B_{\mu\nu}$ &
    $O_{\underset{ij}{dW}}$ & $\bigl(\bar q_i \sigma^{\mu\nu} d_j \bigr) \tau^I \varphi W_{\mu\nu}^I$ & $O_{\underset{ij}{dG}}$ & $\bigl(\bar q_i \sigma^{\mu\nu} T^A d_j \bigr) \varphi G_{\mu\nu}^A$ \\
    \hline
    \multicolumn{6}{|c|}{{Semileptonic Four-Fermion}} \\
    \hline
    $O_{\underset{\alpha\beta ij}{lq}}^{(1)}$ & $\bigl(\bar l_\alpha \gamma_{\mu} l_\beta \bigr)\bigl(\bar q_i \gamma^{\mu} q_j \bigr)$ &
    $O_{\underset{\alpha\beta ij}{lq}}^{(3)}$ & $\bigl(\bar l_\alpha \gamma_{\mu} \tau^I l_\beta \bigr)\bigl(\bar q_i \gamma^{\mu} \tau^I q_j \bigr)$ & $O_{\underset{\alpha\beta ij}{qe}}$ & $\bigl(\bar q_i \gamma_{\mu} q_j \bigr)\bigl(\bar e_\alpha \gamma^{\mu} e_\beta \bigr)$ \\
    $O_{\underset{\alpha\beta ij}{lu}}$ & $\bigl(\bar l_\alpha \gamma_{\mu} l_\beta \bigr)\bigl(\bar u_i \gamma^{\mu} u_j \bigr)$ & $O_{\underset{\alpha\beta ij}{ld}}$ & $\bigl(\bar l_\alpha \gamma_{\mu} l_\beta \bigr)\bigl(\bar d_i \gamma^{\mu} d_j \bigr)$ & $O_{\underset{\alpha\beta ij}{eu}}$ & $\bigl(\bar e_\alpha \gamma_{\mu} e_\beta \bigr)\bigl(\bar u_i \gamma^{\mu} u_j \bigr)$ \\
    $O_{\underset{\alpha\beta ij}{ed}}$ & $\bigl(\bar e_\alpha \gamma_{\mu} e_\beta \bigr)\bigl(\bar d_i \gamma^{\mu} d_j \bigr)$ & $O_{\underset{\alpha\beta ij}{ledq}} $ & $\bigl(\bar l^k_\alpha e_\beta \bigr)\bigl(\bar d_i q^k_j \bigr)$ & $O_{\underset{\alpha\beta ij}{lequ}}^{(1)}$ & $\bigl(\bar l^k_\alpha e_\beta \bigr)\epsilon_{km}\bigl(\bar q^m_i u_j \bigr)$ \\
    $O_{\underset{\alpha\beta ij}{lequ}}^{(3)}$ & $\bigl(\bar l^k_\alpha \sigma_{\mu\nu} e_\beta \bigr)\epsilon_{km}\bigl(\bar q^m_i u_j \bigr)$ & & & & \\
    \hline
  \end{tabular}
  \caption{The operators considered in this work, see text for details.}
  \label{tab:operators}
\end{table}

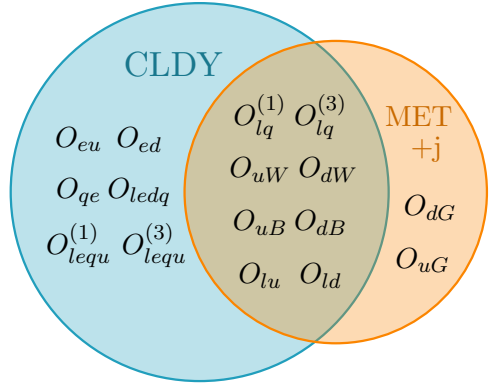
\begin{figure}[ht]
  \xdefinecolor{dRed}{RGB}{153, 0, 0}
\xdefinecolor{dOrange}{RGB}{251, 133, 0}
\xdefinecolor{dBlue}{RGB}{33, 158, 188}
\xdefinecolor{dOrangeDarker}{RGB}{201, 106, 0}
\xdefinecolor{dBlueDarker}{RGB}{25, 119, 141}
\xdefinecolor{dGrey}{RGB}{100, 100, 100}
\xdefinecolor{dBlack}{RGB}{0, 0, 0}

\begin{tikzpicture}
    \draw[color=dBlue, fill=dBlue, fill opacity=0.3, line width=0.3mm] (0,0) circle (2.5cm);
    \draw[color=dOrange, fill=dOrange, fill opacity=0.3, line width=0.3mm] (1.8,0) circle (2.0cm);

    \node[text=dBlueDarker, scale=1.2] at (-0.35,1.7) {CLDY};
    \node[text=dOrangeDarker] at (2.9,1.0) {MET};
    \node[text=dOrangeDarker] at (3,0.57) {+j};

    \node[text=] at (3.05,-0.2) {$O_{dG}$};
    \node[text=dBlack] at (2.95,-0.9) {$O_{uG}$};

    \node[text=dBlack] at (0.8,1.0) {$O_{lq}^{(1)}$};
    \node[text=dBlack] at (1.6,1.0) {$O_{lq}^{(3)}$};
    \node[text=dBlack] at (0.8,0.3) {$O_{uW}$};
    \node[text=dBlack] at (1.7,0.3) {$O_{dW}$};
    \node[text=dBlack] at (0.8,-0.4) {$O_{uB}$};
    \node[text=dBlack] at (1.6,-0.4) {$O_{dB}$};
    \node[text=dBlack] at (0.8,-1.1) {$O_{lu}$};
    \node[text=dBlack] at (1.6,-1.1) {$O_{ld}$};

    \node[text=dBlack] at (-1.6,0.7) {$O_{eu}$};
    \node[text=dBlack] at (-0.8,0.7) {$O_{ed}$};
    \node[text=dBlack] at (-1.6,0.0) {$O_{qe}$};
    \node[text=dBlack] at (-0.8,0.0) {$O_{ledq}$};

    \node[text=dBlack] at (-1.6,-0.7) {$O_{lequ}^{(1)}$};
    \node[text=dBlack] at (-0.6,-0.7) {$O_{lequ}^{(3)}$};

\end{tikzpicture}
  \caption{Sensitivities of the CLDY (blue) and MET+j (orange) processes to the dimension-six SMEFT operators considered in this work.}
  \label{fig:operators}
\end{figure}

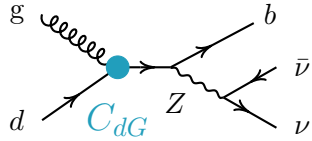
\begin{figure}[ht]
  \xdefinecolor{dRed}{RGB}{153, 0, 0}
\xdefinecolor{dOrange}{RGB}{251, 133, 0}
\xdefinecolor{dBlue}{RGB}{33, 158, 188}
\xdefinecolor{dGrey}{RGB}{100, 100, 100}

\begin{tikzpicture}[scale=1] \begin{feynman}

  \vertex (ww32);
  \vertex [right=0.3cm of ww32] (ww33);
  \vertex [right=0.7cm of ww32] (ww42);

  \vertex[left=1cm of ww32] (ml22);
  \vertex[above=0.7cm of ml22] (gg12);
  \vertex[left=0.1cm of gg12] (g12) {g};
  \vertex[below=0.7cm of ml22] (qq12);
  \vertex[left=0.1cm of qq12] (q12) {\(d\)};

  \vertex[right=1cm of ww42] (mr22);
  \vertex[above=0.7cm of mr22] (qq22);
  \vertex[right=0.1cm of qq22] (q22) {\(b\)};
  \vertex[below=0.4cm of mr22] (h12);
  \vertex[left=0.3cm of h12] (Z2);

  \vertex[right=0.7cm of Z2] (mnu2);
  \vertex[above=0.4cm of mnu2] (nu12);
  \vertex[below=0.4cm of mnu2] (nu22);
  \vertex[right=0.1cm of nu12] (nubar2) {$\bar \nu$};
  \vertex[right=0.1cm of nu22] (nu42) {$\nu$};

  \diagram* {

    (ww32) -- [thick] (ww33) -- [thick,fermion] (ww42) -- [thick, fermion] (q22),
    (gg12) -- [thick,gluon] (ww32),
    (qq12) -- [thick, fermion] (ww32),
    (ww42) -- [thick, boson, edge label'=\({Z}\)] (Z2),

    (nu12) -- [thick,fermion] (Z2) -- [thick, fermion] (nu22),
      };

      \draw (0.0,-0.7) node[ dBlue, font=\large] {$ C_{dG}$};

      \draw[fill=dBlue, dBlue] (0.0,-0.0) circle (0.15) node[right=5pt] {};

\end{feynman} \end{tikzpicture}
  \caption{Contribution of the gluon dipole operator $O_{dG}$ (blue blob) to the MET+j process.}
  \label{fig:gluonDipole}
\end{figure}

The sensitivities of the CLDY (blue) and MET+j (orange) processes to the SMEFT coefficients are illustrated in Fig.~\ref{fig:operators}. While both processes are sensitive to the quark dipole operators $O_{qW}$ and $O_{qB}$, the dineutrino process is only sensitive to semileptonic four-fermion operators involving left-handed leptons~\footnote{Right-handed neutrinos are not considered in this analysis, they can, however, be probed with MET+j~\cite{Hiller:2024vtr}.}. In contrast, CLDY also probes right-handed lepton couplings, $C_{eu}, C_{ed}$ and $C_{qe}$, in addition to the scalar and tensor coefficients $C_{ledq}$, $C_{lequ}^{(1)}$, and $C_{lequ}^{(3)}$. On the other hand, the MET+j process is sensitive to the gluon dipole operators $O_{qG}$, as shown in Fig.~\ref{fig:gluonDipole}, whereas these operators do not contribute to the CLDY process at leading order.

\subsection{Quark flavor bases}
\label{sec:mass_basis}

The aim of this works is to explore synergies  between the two types of Drell-Yan, seemingly different in charged dileptons and MET plus jet, and also combined with rare decays of hadrons.
While flavor-diagonal contributions in particular for light quarks would be subject to stronger constraints from Drell-Yan data, to make contact with the rare decays, here we focus on
quark flavor specific operators that induce FCNC processes. In a given model of flavor, or framework, e.g.~\cite{Bissmann:2020mfi, Grunwald:2023nli} of course, a joint analysis becomes feasible, and would allow for tests of the flavor model.

To confront the SMEFT to experimental data, fermion fields in the operators need to be rotated from the gauge to the mass basis. 
This is immaterial for operators involving only singlet quarks, as their rotation is unphysical and can be absorbed by redefining the corresponding WCs. 
The same is true in our analysis concentrating on neutral currents for
operators involving chirality-flipping quark currents, such as the dipoles, scalar and tensor operators. With additional
charged current observables, however, the impact of flavor rotations cannot be fully absorbed in the WCs.
For operators with doublet quark currents, $O_{lq}^{(1)}$, $O_{lq}^{(3)}$ and $O_{qe}$ the rotation has a physical effect.
We take this into account by performing the global fit in two extreme cases,
up- and down-alignment, meaning that in the former the flavor rotations responsible for the CKM-matrix reside entirely in the down-sector, while in the latter in the up-sector~\cite{Allwicher:2023shc}.
We illustrate this for $O_{lq}^{(1)}, O_{lq}^{(3)}$, whose contributions expanded in components read
\begin{equation}
  \begin{aligned}
    \label{eqn:Lagr}
    \mathcal{L} \propto \, &C^{(1)}_{\underset{\alpha \beta ij}{lq}} \bigl( \bar q_i \gamma^{\mu} q_j \bigr)\bigl( \bar l_{\alpha} \gamma_{\mu} l_{\beta} \bigr) + C^{(3)}_{\underset{\alpha \beta ij}{lq}} \bigl( \bar q_i \gamma^{\mu}\tau^I q_j \bigr) \bigl(\bar l_{\alpha} \gamma_{\mu} \tau^I l_{\beta} \bigr) \\
    =\, &C^{+}_{\underset{\alpha \beta ij}{lq}} \Bigl( \bigl(\bar u^L_i \gamma^{\mu} u^L_j \bigr) \bigl(\bar \nu_{\alpha} \gamma_{\mu} \nu_{\beta} \bigr) + \bigl( \bar d^L_i \gamma^{\mu} d^L_j \bigr) \bigl( \bar \ell_{\alpha} \gamma_{\mu} \ell_{\beta} \bigr) \Bigr) \\
    + \,&C^{-}_{\underset{\alpha \beta ij}{lq}} \Bigl( \bigl( \bar d^L_i \gamma^{\mu} d^L_j \bigr) \bigl( \bar \nu_{\alpha} \gamma_{\mu} \nu_{\beta}\bigr) + \bigl( \bar u^L_i \gamma^{\mu} u^L_j \bigr) \bigl( \bar \ell_{\alpha} \gamma_{\mu} \ell_{\beta} \bigr) \Bigr) \, ,
  \end{aligned}
\end{equation}
in the gauge basis, where 
\begin{equation}
  C^{\pm}_{lq} = C^{(1)} _{lq} \pm C^{(3)}_{lq} \,.
  \label{eqn:Cpm}
\end{equation}
The quark fields are rotated to the mass basis with unitary rotations $U^{u,d}_{L/R}$
\begin{equation}
  \label{eqn:Rot}
  u_{L/R} = U^{u}_{L/R} u^{\prime}_{L/R} \,, ~~~
  d_{L/R} = U^{d}_{L/R} d^{\prime}_{L/R} \,,~~~ V = U^{u\dagger}_L U_L^d \, , 
\end{equation}
where the primed (unprimed) fields denote the mass (gauge) eigenstates. $V$ is the CKM matrix.
For the example of $C^{+}_{lq}$, the rotation to mass eigenstates gives
\begin{equation}
  C^{+}_{\underset{\alpha \beta ij}{lq}} \left( U^{*u}_{\underset{ik}{L}} U^{u}_{\underset{jm}{L}} \bar u^{\prime}_k \gamma^{\mu} u^{\prime}_m \bar \nu_{\alpha} \gamma_{\mu} \nu_{\beta} + U^{*d}_{\underset{ik}{L}} U^{d}_{\underset{jm}{L}} \bar d^{\prime}_k \gamma^{\mu} d^{\prime}_m \bar \ell_{\alpha} \gamma_{\mu} \ell_{\beta} \right) \,.
\end{equation}
In up-alignment (down-alignment), the $U_L^u$ rotation matrices ($U_L^d$ matrices) are absorbed, i.e.
$ \hat C^{+, \text{up}}_{\underset{\alpha \beta ij}{lq}} = U^{*u}_{\underset{ik}{L}}U^{u}_{\underset{jm}{L}} C^{+}_{\underset{\alpha \beta km}{lq}}$ and
 $\hat C^{+, \text{down}}_{\underset{\alpha \beta ij}{lq}} = U^{*d}_{\underset{ik}{L}}U^{d}_{\underset{jm}{L}} C^{+}_{\underset{\alpha \beta km}{lq}}$.
In up-alignment, we then obtain
\begin{equation}
  \begin{aligned}
    \label{eqn:Lagr_up}
    \mathcal{L} & \propto \hat C^{+, \text{up}}_{\underset{\alpha \beta ij}{lq}} \left(\bar u^{\prime}_i \gamma^{\mu} u^{\prime}_j \bar \nu_{\alpha} \gamma_{\mu} \nu_{\beta} + V^*_{ik}V_{jm} \bar d^{\prime}_k \gamma^{\mu} d^{\prime}_m \bar \ell_{\alpha} \gamma_{\mu} \ell_{\beta} \right) \\
    &+ \hat C^{-, \text{up}}_{\underset{\alpha \beta ij}{lq}} \left(V^*_{ik}V_{jm} \bar d^{\prime}_k \gamma^{\mu} d^{\prime}_m \bar \nu_{\alpha} \gamma_{\mu} \nu_{\beta} +\bar u^{\prime}_i \gamma^{\mu} u^{\prime}_j \bar \ell_{\alpha} \gamma_{\mu} \ell_{\beta} \right) \, ,
  \end{aligned}
\end{equation}
whereas in down-alignment, the Lagrangian reads 
\begin{equation}
  \begin{aligned}
    \label{eqn:Lagr_down}
    \mathcal{L} & \propto \hat C^{+, \text{down}}_{\underset{\alpha \beta ij}{lq}} \left(V_{ki}V^*_{mj} \bar u^{\prime}_k \gamma^{\mu} u^{ \prime}_m \bar \nu_{\alpha} \gamma_{\mu} \nu_{\beta} + \bar d^{\prime}_i \gamma^{\mu} d^{\prime}_j \bar \ell_{\alpha} \gamma_{\mu} \ell_{\beta} \right) \, \\
    &+ \hat C^{-, \text{down}}_{\underset{\alpha \beta ij}{lq}} \left( \bar d^{\prime}_i \gamma^{\mu} d^{\prime}_j  \bar \nu_{\alpha} \gamma_{\mu} \nu_{\beta} + V_{ki}V^*_{mj} \bar u^{\prime}_k \gamma^{\mu} u^{ \prime}_m \bar \ell_{\alpha} \gamma_{\mu} \ell_{\beta} \right) \,.
  \end{aligned}
\end{equation}
We learn that additional contributions, FCNC and flavor diagonal ones whose sizes are driven by CKM-hierarchies, arise in each case (\eqref{eqn:Lagr_up}), (\eqref{eqn:Lagr_down}).
Note that for $i \neq j$ the mixing-induced flavor diagonal contributions ($k=m$), which interfere with the SM and are CKM-suppressed.
We include the full expressions (\ref{eqn:Lagr_up}), (\ref{eqn:Lagr_down}) in our analysis.
To avoid clutter in the remainder of this work we omit hats on Wilson coefficients and tacitly assume that they correspond to operators with quarks in the mass basis.

\subsection{Lepton flavor patterns}
\label{sec:leptons}

We analyze different scenarios for the lepton-flavor structure of the SMEFT coefficients, a lepton flavor universal (LU), a lepton-flavor violating (LFV) and a democratic one.
We consider also lepton-flavor-specific couplings, where each coefficient ${C}_{\alpha\beta ij}$ for each $\alpha$ and $\beta$ is treated as an independent parameter. The CLDY process is measured lepton-flavor specifically, allowing for the resolution of individual flavor contributions. In contrast, the MET+j process constrains the incoherent sum of lepton flavors, as the individual contributions are experimentally not resolved. Despite this limitation, it is still possible to constrain all coefficients simultaneously, as there is no interference between different flavor contributions in the effective coefficients. Hence, the bounds on the individual lepton-flavor coefficients are at best as strong as those on the incoherent sum.
For instance, 
\begin{equation}
| C_{11ij}|^2 \leq \sum_{\alpha,\beta} | C_{\alpha\beta ij}|^2 \,,
\label{eqn:sum_lepton_flavor}
\end{equation}
holds for the dielectron coupling $ C_{11ij}$ for any quark indices $i,j$.

For flavor off-diagonal couplings $\alpha \neq \beta$, an additional factor of $\sqrt{2}$ arises for hermitian operators due to \( C_{\alpha\beta ij} = C_{\beta\alpha ij}^* \). For example, 
switching on $\alpha,\beta=1,2$ couplings implies
\begin{equation}
\sum_{\alpha,\beta} | C_{\alpha\beta ij}|^2 = | C_{12ij}|^2 + | C_{21ij}|^2 \leq c^2 \,,
\end{equation}
and
\begin{equation}
| C_{12ij}| \leq \frac{c}{\sqrt{2}} \,.
\end{equation}

We also consider lepton-flavor universal patterns. Here, $ C_{\alpha\beta ij} = \delta_{\alpha\beta} C^{\text{LU}}_{ij}$, i.e.
diagonal elements are assumed to be equal, while off-diagonal, flavor violating elements vanish\footnote{In MFV, the scalar and tensor contributions are suppressed by small lepton Yukawas.}. The sum over the lepton flavors then simplifies to
\begin{equation}
  \sum_{\alpha,\beta} | C_{\alpha\beta ij}|^2 = 3| C^{\text{LU}}_{ij}|^2 \leq c^2 \,,
\end{equation} 
implying that the LU bound is by a factor $\sqrt{3}$ stronger than the lepton-flavor-specific one
\begin{equation}
  | C^{\text{LU}}_{ij}| \leq \frac{c}{\sqrt{3}} \,.
\end{equation}

This scenario can be extended by allowing for an additional LFV contribution that is universal among the non-diagonal couplings, i.e. $ C_{\alpha\beta ij} = \delta_{\alpha\beta}  C^{\text{LU}}_{ij} + \delta_{\alpha \neq \beta}  C^{\text{LFV}}_{ij}$. This pattern is motivated by scenarios with different scales for lepton-flavor conserving and lepton-flavor violating new physics. For the sum over the lepton flavors in the MET+j observables, we find
\begin{equation}
 \sum_{\alpha,\beta} | C_{\alpha\beta ij}|^2 =3 | C^{\text{LU}}_{ij}|^2 + 6 | C^{\text{LFV}}_{ij} |^2\leq c^2 \,,
\end{equation}
implying that the bounds from the MET+j process are by a factor of $\sqrt{2}$ stronger for the LFV contributions than for the lepton-flavor conserving ones.

In addition we consider a democratic pattern, where all lepton-flavor couplings are assumed to be equal, i.e. $ C_{\alpha\beta ij} = C^{\text{dem}}_{ij}$. Consequently, the sum over the lepton flavors in the MET+j process scales as 
\begin{equation}
  \sum_{\alpha,\beta} | C_{\alpha\beta ij}|^2 = 9 | C^{\text{dem}}_{ij}|^2 \leq c^2 \,,
\end{equation}
leading to a factor of 3 improvement in the bounds on the democratic couplings compared to the lepton-flavor specific ones.

\subsection{Parton level cross sections}
\label{sec:cross_sections}

Both types of Drell-Yan processes are sensitive to specific combinations of WCs, which can be determined at parton level and directly mapped onto the relevant collider observables. 
Generically SMEFT induces terms quadratic and linear in the WCs. However for vanishing fermion masses, only linear terms with vectorial operators with diagonal quark and lepton flavor indices remain. 
In the high energy limit $\hat s \gg M_Z^2$, the parton level cross section $\hat \sigma$ of the CLDY process can be parameterized as~\cite{Allwicher:2022gkm}
\begin{equation}
    \begin{aligned}
    \hat \sigma\left( q_i \bar q_j \rightarrow \ell_{\alpha}\ell_{\beta}\right) &= \hat \sigma_{\text{SM}}\left( q \bar q \rightarrow \ell\ell\right) \delta_{ij} \delta_{\alpha \beta}  
    +  C^{\ell \ell}_{\underset{\alpha\beta ij}{\text{4F},\text{lin}}} \hat\sigma_{\text{4F}}^{\text{lin}}\left( q \bar q \rightarrow \ell\ell\right) \delta_{ij} \delta_{\alpha \beta}  \\
    &+ \left(C^{\ell \ell}_{\underset{\alpha\beta ij}{\text{4F},\text{quad}}}\right)^2 \hat\sigma_{\text{4F}}^{\text{quad}}\left( q \bar q \rightarrow \ell \ell\right) 
   + \left(C^{\ell \ell}_{\underset{ij}{\text{EW}}}\right)^2 \hat \sigma_{\text{EW}}\left( q \bar q \rightarrow \ell \ell\right) \delta_{\alpha \beta }\,,
  \end{aligned}
  \label{eqn:xsec_ll}
\end{equation}
where the effective coefficients are given by
\begin{align}
C^{\ell \ell}_{\underset{\alpha\alpha ii}{\text{4F},\text{lin}}} = & 
\begin{cases}
  -\left( C^{-}_{\underset{\alpha \alpha ii}{lq}} \left( 1 +3 \frac{c_W^2}{s_W^2}\right) + 4C_{\underset{\alpha \alpha ii}{lu}} +8C_{\underset{\alpha \alpha ii}{eu}} +2 C_{\underset{\alpha \alpha ii}{qe}} \right) \quad \text{(up-type quarks)} \,, \\
  \left( C^{+}_{\underset{\alpha \alpha ii}{lq}} \left( -1 +3 \frac{c_W^2}{s_W^2}\right) + 2C_{\underset{\alpha \alpha ii}{ld}} +4C_{\underset{\alpha \alpha ii}{ed}} -2 C_{\underset{\alpha \alpha ii}{qe}} \right) \quad \text{(down-type quarks)} \,, \\
\end{cases}
\end{align}
\begin{align}
  \left(C^{\ell \ell}_{\underset{\alpha\beta ij}{\text{4F},\text{quad}}}\right)^2 = &
  \begin{cases}
     |C_{\underset{\alpha \beta ij}{eu}}|^2 + |C_{\underset{\alpha \beta ij}{qe}}|^2 + |C_{\underset{\alpha \beta ij}{lu}}|^2 + |C^{-}_{\underset{\alpha \beta ij}{lq}}|^2 +\frac{3}{4}|C^{(1)}_{\underset{\alpha \beta ij}{lequ}}|^2 + 4|C^{(3)}_{\underset{\alpha \beta ij}{lequ}}|^2 \quad \text{(up-type quarks)} \,, \\
      |C_{\underset{\alpha \beta ij}{ed}}|^2 + |C_{\underset{\alpha \beta ij}{qe}}|^2 + |C_{\underset{\alpha \beta ij}{ld}}|^2 + |C^{+}_{\underset{\alpha \beta ij}{lq}}|^2 + \frac{3}{4}|C_{\underset{\alpha \beta ij}{ledq}}|^2 \quad \text{(down-type quarks)} \,, \\
  \end{cases}
  \label{eqn:C_4F_ll}
\end{align}
and 
\begin{fleqn}[\parindent]
  \begin{align}
    \raisebox{0.8\height}{$\left(C^{\ell \ell}_{\underset{ij}{\text{EW}}}\right)^2= \ $}&
    \begin{aligned}
    &\ \frac{1 - 4 s_W^2 + 8 s_W^4}{4c_W^2 s_W^2}\left( |C_{\underset{ij}{q\gamma}}|^2 +|C_{\underset{ji}{q\gamma}}|^2\right) + \left(|C_{\underset{ij}{qZ}} |^2 +|C_{\underset{ji}{qZ}} |^2\right) \\
    &-2c_W s_W\frac{1- 4 s_W^2}{1 - 4 s_W^2 + 9 s_W^4} \text{Re}\left\{ C_{\underset{ij}{q\gamma}} C_{\underset{ij}{qZ}}^* + C_{\underset{ji}{q\gamma}} C_{\underset{ji}{qZ}}^* \right\} \,, \hfill
    \end{aligned}
  \label{eqn:C_EW_ll}
  \end{align}
\end{fleqn}
where $c_W = \cos \theta_W , s_W = \sin \theta_W$ with the weak mixing angle $\theta_W$. The effective coefficients are thereby defined as
$C^\pm_{lq}$ in Eq.(\ref{eqn:Cpm})
and 
\begin{align}
    C_{\underset{ij}{q\gamma}} &=
    \begin{cases}
      s_W C_{\underset{ij}{uW}} + c_W C_{\underset{ij}{uB}} &\quad \text{(up-type quarks)} \,, \\
      -s_W C_{\underset{ij}{dW}} + c_W C_{\underset{ij}{dB}} &\quad \text{(down-type quarks)} \,,
    \end{cases}\label{eqn:CqGamma} \\
    C_{\underset{ij}{qZ}} &=
    \begin{cases}
      c_W C_{\underset{ij}{uW}} - s_W C_{\underset{ij}{uB}} &\quad \ \text{(up-type quarks)}\,, \\
     - c_W C_{\underset{ij}{dW}} - s_W C_{\underset{ij}{dB}} &\quad \ \text{(down-type quarks)} \,.
    \end{cases} 
    \label{eqn:CqZ} 
  \end{align}

For the MET+j process, the $g q$-channel provides the dominant contribution to the $\emiss$-spectrum, as discussed in detail in Ref.~\cite{Hiller:2024vtr}. The differential 
parton level cross section can be parameterized as
\begin{equation}
  \begin{aligned}
    \frac{\mathrm{d}\hat \sigma\left( q_i g \rightarrow \nu \bar \nu q_j\right) }{\mathrm{d} E_T^{\text{miss}}} &= \frac{\mathrm{d}\hat \sigma_{SM}\left( q g \rightarrow \nu \bar \nu q\right) }{\mathrm{d} E_T^{\text{miss}}} \delta_{ij} \\
    &+(C^{\nu \nu}_{\underset{ij}{\text{4F},\text{lin}}})\frac{\mathrm{d}\hat \sigma_{4F}\left( q g \rightarrow \nu \bar \nu q\right) }{\mathrm{d} E_T^{\text{miss}}} \delta_{ij} +(C^{\nu \nu}_{\underset{ij}{\text{4F},\text{quad}}})^2\frac{\mathrm{d}\hat \sigma_{4F}\left( q g \rightarrow \nu \bar \nu q\right) }{\mathrm{d} E_T^{\text{miss}}} \\
    &+(C^{{\nu \nu}}_{\underset{ij}{\text{EW}}})^2 \frac{\mathrm{d}\hat \sigma_{EW}\left( q g \rightarrow \nu \bar \nu q\right) }{\mathrm{d} E_T^{\text{miss}}} +(C^{\nu \nu}_{\underset{ij}{\text{G}}})^2 \frac{\mathrm{d}\hat \sigma_{G}\left( q g \rightarrow \nu \bar \nu q\right) }{\mathrm{d} E_T^{\text{miss}}} \,,
  \end{aligned}
  \label{eqn:xsec_nunu}
\end{equation}
where we introduce the effective coefficients
\begin{align}
  &C^{\nu \nu}_{\underset{ii}{\text{4F},\text{lin}}} = 
  \begin{cases}
    \sum_{\alpha,\beta} \epsilon_L^u C^{+}_{\underset{\alpha \beta ii}{lq}}+ \epsilon_R^u C_{\underset{\alpha \beta ii}{lu} } & \text{(up-type quarks)} \,,\\
    \sum_{\alpha, \beta} \epsilon_L^d C^{-}_{\underset{\alpha \beta ii}{lq}}+ \epsilon_R^d C_{\underset{\alpha \beta ii}{ld}} & \text{(down-type quarks)} \,,
  \end{cases}\label{eqn:C_4F_nunu_lin} \\
  &(C^{\nu \nu}_{\underset{ij}{\text{4F},\text{quad}}})^2 =
  \begin{cases}
    \sum_{\alpha,\beta} |C^{+}_{\underset{\alpha \beta ij}{lq}}|^2+ |C_{\underset{\alpha \beta ij}{lu}}|^2 & \text{(up-type quarks)} \,,\\
    \sum_{\alpha, \beta} |C^{-}_{\underset{\alpha \beta ij}{lq}}|^2+ |C_{\underset{\alpha \beta ij}{ld}}|^2 & \text{(down-type quarks)} \,,
  \end{cases}\label{eqn:C_4F_nunu} \\
  &(C^{\nu \nu}_{\underset{ij}{\text{EW}}})^2 = |C_{\underset{ij}{qZ}}|^2 + |C_{\underset{ji}{qZ}}|^2 \,,
  \label{eqn:C_EW_nunu}\\
  &(C^{\nu \nu}_{\underset{ij}{\text{G}}})^2 = |C_{\underset{ij}{qG}}|^2 + |C_{\underset{ji}{qG}}|^2 \,,
  \label{eqn:C_G_nunu}
\end{align} 
with $q=u,d$ and $\epsilon_X^{q}$ with $X = L,R$ is defined in Eq.\eqref{eqn:Z_couplings_SM}. 
Analogous parameterizations for the $q\bar q$-channel are given in Ref.~\cite{Hiller:2024vtr}. These channels exhibit the same dependence on the SMEFT coefficients, whereas they differ with regard to the kinematic terms.

Here, all WCs are given in the gauge basis. The non-trivial rotations to the mass basis are given by 
\begin{align}
  \label{eqn:C_lq_plus_mass}
  &C^{+}_{\underset{\alpha \beta ij}{lq}} \rightarrow \begin{cases}
    C^{+}_{\underset{\alpha \beta km}{lq}} V_{ki}^{*} V_{mj} &\quad \text{$\ell\ell$, up-alignment} \,, \\
    C^{+}_{\underset{\alpha \beta km}{lq}} V_{ik} V_{jm}^{*} &\quad \text{$\nu\nu$, down-alignment} \,, \\
  \end{cases} \\
  \label{eqn:C_lq_minus_mass}
  &C^{-}_{\underset{\alpha \beta ij}{lq}} \rightarrow
   \begin{cases}
    C^{-}_{\underset{\alpha \beta km}{lq}} V_{ki}^{*} V_{mj} &\quad \text{$\nu\nu$, up-alignment} \,, \\
    C^{-}_{\underset{\alpha \beta km}{lq}} V_{ik} V_{jm}^{*} &\quad \text{$\ell\ell$, down-alignment} \,, 
  \end{cases} \\
  \label{eqn:C_qe_mass}
  &C_{\underset{\alpha \beta ij}{qe}} \rightarrow \begin{cases}
    C_{\underset{\alpha \beta km}{qe}} V_{ki}^{*} V_{mj} &\quad \text{$\ell\ell$, down-type quarks, up-alignment} \,, \\
    C_{\underset{\alpha \beta km}{qe}} V_{ik} V_{jm}^{*} &\quad \text{$\ell\ell$, up-type quarks, down-alignment} \,. \\
  \end{cases} \\
\end{align}
More details on the alignment are given in Sec.~\ref{sec:mass_basis}.

Let us stress the inclusiveness of the analysis regarding multiple operators, lepton and quark flavors:
As the four-fermion operators comprise different chiralities of the fermions, they contribute incoherently to the cross sections, e.g. (\ref{eqn:C_4F_ll}). Similarly, the effective coefficients defined in Eq.~\eqref{eqn:C_4F_nunu} constrain the incoherent sum of the lepton flavors. 
In addition, the hadron level cross sections, see e.g.~Ref.~\cite{Grunwald:2023nli} for CLDY and~\cite{Hiller:2024vtr} for MET+j for details,
are computed inclusively over the initial quark flavors. 
In all cases, upper limits on the magnitude of individual coefficients can be obtained because 'the sum of squares is an upper limit on each
square'. On the other hand, assuming patterns of contributions, for instance assuming a concrete BSM model that induces only a subset of operators perhaps even correlated
by a few model parameters, or a given lepton-flavor structure, see Sec.~\ref{sec:leptons}, allows for stronger 
limits due to correlations.

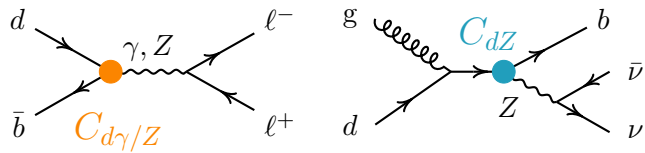
\begin{figure}[h]
  \xdefinecolor{dRed}{RGB}{153, 0, 0}
\xdefinecolor{dOrange}{RGB}{251, 133, 0}
\xdefinecolor{dBlue}{RGB}{33, 158, 188}
\xdefinecolor{dGrey}{RGB}{100, 100, 100}

\begin{tikzpicture}[scale=1] \begin{feynman}

  \vertex (ww1);
  \vertex [left=0.7cm of ww1] (ml);
  \vertex [right=1.cm of ww1] (gam);
  \vertex [right=0.7cm of gam] (mr);
  \vertex [above=0.7cm of ml] (u1);
  \vertex [left=0.3cm of u1] (u) {\(d\)};
  \vertex [below=0.7cm of ml] (u2);
  \vertex [left=0.3cm of u2] (u3) {\(\bar b\)};
  \vertex [above=0.7cm of mr] (l1);
  \vertex [right=0.2cm of l1] (l) {\(\ell^-\)};
  \vertex [below=0.7cm of mr] (l2);
  \vertex [right=0.2cm of l2] (l3) {\(\ell^+\)};

  \vertex [right=4.5cm of ww1] (ww3);
  \vertex [right=0.3cm of ww3] (ww13);
  \vertex [right=0.7cm of ww3] (ww4);

  \vertex[left=1cm of ww3] (ml2);
  \vertex[above=0.7cm of ml2] (gg1);
  \vertex[left=0.1cm of gg1] (g1) {g};
  \vertex[below=0.7cm of ml2] (qq1);
  \vertex[left=0.1cm of qq1] (q1) {\(d\)};

  \vertex[right=1cm of ww4] (mr2);
  \vertex[above=0.7cm of mr2] (qq2);
  \vertex[right=0.1cm of qq2] (q2) {\(b\)};
  \vertex[below=0.4cm of mr2] (h1);
  \vertex[left=0.3cm of h1] (Z);

  \vertex[right=0.7cm of Z] (mnu);
  \vertex[above=0.4cm of mnu] (nu1);
  \vertex[below=0.4cm of mnu] (nu2);
  \vertex[right=0.1cm of nu1] (nubar) {$\bar \nu$};
  \vertex[right=0.1cm of nu2] (nu) {$\nu$};

  \diagram* {
    (u) -- [thick,fermion] (ww1),
    (ww1) -- [thick,fermion] (u3),
    (ww1) -- [thick, boson, edge label=\({\gamma,Z}\)] (gam),
    (l3) -- [thick,fermion] (gam),
    (gam) -- [thick,fermion] (l),

    (ww3) -- [thick] (ww13) -- [thick,fermion] (ww4) -- [thick, fermion] (q2),
    (gg1) -- [thick,gluon] (ww3),
    (qq1) -- [thick, fermion] (ww3),
    (ww4) -- [thick, boson, edge label'=\({Z}\)] (Z),

    (nu1) -- [thick,fermion] (Z) -- [thick, fermion] (nu2),
      };

      \draw[fill=dOrange, dOrange] (0,-0.0) circle (0.15) node[right=5pt] {};
      \draw[fill=dBlue, dBlue] (5.2,0.0) circle (0.15) node[right=5pt] {}; 
    
      \draw (0.1,-0.8) node[ dOrange, font=\large] {$ C_{d \gamma/Z}$};
      \draw (5.0,0.5) node[ dBlue, font=\large] {$ C_{dZ}$};

  \end{feynman} \end{tikzpicture}

  
  \caption{Synergies between CLDY and MET+j for the example of the $db$ coupling for the electroweak dipole operators.
  SMEFT contributions to the CLDY process are highlighted with orange vertices, while MET+j vertices are shown in blue.}
  \label{fig:synergies_dipole}
\end{figure}

The effective coefficients associated with the electroweak dipole operators, defined in Eq.~\eqref{eqn:C_EW_ll} for CLDY and in Eq.~\eqref{eqn:C_EW_nunu} for MET+j, together enable an enhanced resolution of the individual components $(C_{u/dZ}, C_{u/d\gamma})$ or the $SU(2)_L$ and hypercharge directions $(C_{u/dW}, C_{u/dB})$. While CLDY simultaneously probes $Z$-boson and photon couplings, the dineutrino process is only sensitive to the $Z$-coefficient. 
This is illustrated for the example of the $db$ coupling in Fig.~\ref{fig:synergies_dipole}.

\section{LHC-Data and fit setup}
\label{sec:fit_data}

We outline the strategy to obtain limits on SMEFT operators from LHC DY measurements. We begin by specifying the processes and data sets used in our analysis, comprising of the CLDY and MET+j channels. We then describe our Monte Carlo (MC) simulation setup and validation procedures, as well as detail the Bayesian framework employed to extract the constraints on the WCs.

\subsection{Data sets and observables}
\label{sec:data_sets}

For the CLDY process, we consider analyses of the lepton-flavor conserving process ${p p \to \ell_{\alpha}^+ \ell_{\alpha}^-}$,
as well as searches for lepton-flavor violating processes ${p p \to \ell_{\alpha}^+ \ell_{\beta}^-}$, with $\alpha \neq \beta$ and $\alpha,\beta = e, \mu, \tau$. For the $p p \to \tau^+ \tau^- $ process, we only use the leptonic channel with veto against $b$-jets, since the correlations with other channels are not provided and it has been shown that these correlations can significantly impact the fit~\cite{Bissmann:2019qcd}.

For the MET+j process, we consider the process ${p p \to \nu \bar{\nu} j}$, where the dineutrino pair is produced in association with high-$p_T$ jets. The data sets used in the analysis are summarized in Tab.~\ref{tab:data_sets}.
\begin{table}[h]
  \centering
  \begin{tabular}{c c c c | c c c c}
    Process & Observable & $\mathcal{L}_{\text{int}}$ & Ref. & Process & Observable & $\mathcal{L}_{\text{int}}$ & Ref. \\
    \toprule
    $p p \to e^+ e^- $ & $\frac{\text{d}\sigma}{\text{d}m_{\ell \ell}}$ & 137 fb$^{-1}$ & \cite{CMS:2021ctt} & $p p \to e \mu, e \tau, \mu \tau $ & $\frac{\text{d}\sigma}{\text{d}m_{\ell \ell^{\prime}}}$ & 139 fb$^{-1}$ & \cite{ATLAS:2020tre} \\
    $p p \to \mu^+ \mu^- $ & $\frac{\text{d}\sigma}{\text{d}m_{\ell \ell}}$ & 140 fb$^{-1}$ & \cite{CMS:2021ctt} & $p p \to \text{MET} +j$ & $\frac{\text{d}\sigma}{\text{d}\emiss}$ & 139 fb$^{-1}$ &\cite{ATLAS:2021kxv} \\
    $p p \to \tau^+ \tau^- $ & $\frac{\text{d}\sigma}{\text{d}m_{\text{T}}^{\text{tot}}}$ & 139 fb$^{-1}$ & \cite{ATLAS:2020zms} & & & & \\
  \end{tabular}
  \caption{Data sets and observables of CLDY and MET+j used in the analysis, together with the corresponding integrated luminosity $\mathcal{L}_{\text{int}}$. All data is taken at $\sqrt{s} = 13$ TeV.}
  \label{tab:data_sets}
\end{table}

\subsection{Simulation of the SMEFT contributions}
\label{sec:simulation}

We simulate the contributions of the SMEFT operators to the CLDY and MET+j processes using {\mg~\cite{Alwall:2014hca}}, with the {\smeftsim~3.0} model~\cite{Brivio:2020onw}. The new physics (NP) scale is set to $\Lambda = 1$ TeV in all simulations. We neglect the running of the SMEFT coefficients, as the effects are 
small for the scales and operators considered in this analysis.

We use the NNPDF4.0~PDF~sets~\cite{NNPDF:2021njg}, employing LO PDF sets for the CLDY processes and NLO PDF sets for the MET+j process, owing to the additional real emission required for tagging. Parton showering and hadronization are handled by \pythia~\cite{Bierlich:2022pfr}, and detector simulation is carried out using \delphes~\cite{deFavereau:2013fsa}. We validate our setup by recasting the SM predictions from the experimental analyses. 

\subsection{Bayesian fit framework}
\label{sec:fit}

To set constraints on the WCs, we employ the \EFTfitter framework~\cite{Castro:2016jjv}, which is based on \bat~\cite{Schulz:2020ebm} and utilizes a Bayesian statistical approach. In this way, we can simultaneously derive credible intervals for all coefficients. We adopt flat priors in the range 
[-10,10] for most coefficients. For those whose 95\% credible intervals extend beyond this range, we set flat priors in the range [-50,50]. We have checked the dependence on the priors by repeating the fits with a Gaussian prior, finding no difference in the results.

We include systematical as well as statistical uncertainties for the background processes, which we assume to be Gaussian distributed. Since we assume the NP signal to be small, we neglect uncertainties on the SMEFT contributions. The sampling of the posterior probability distribution is performed with the robust adaptive Metropolis algorithm~\cite{Vihola:2012}, ensuring an efficient exploration of the parameter space. The 95\% credible limits on the individual SMEFT coefficients are derived by marginalizing over the posterior probability distribution.

\section{Limits \& Correlations}
\label{sec:Results}

We present the new physics reach by CLDY, MET+j and their combination.
In Sec.~\ref{sec:LU_results} we discuss the basic steps and general features of the fit, and present limits assuming lepton-flavor universality.
As there are qualitative differences to the first two quark generations, we discuss fit results involving $b$-quarks separately in Sec.~\ref{sec:b}.
Results in the LFV and the lepton-flavor-specific scenarios, defined in Sec.~\ref{sec:leptons}, are given in Sec.~\ref{sec:LFV_vs_LU} and
Sec.~\ref{sec:LF_specific}, respectively.

\subsection{General features of analysis}
\label{sec:LU_results}

\begin{figure}[h]
  \centering
  \includegraphics[width=0.99\textwidth]{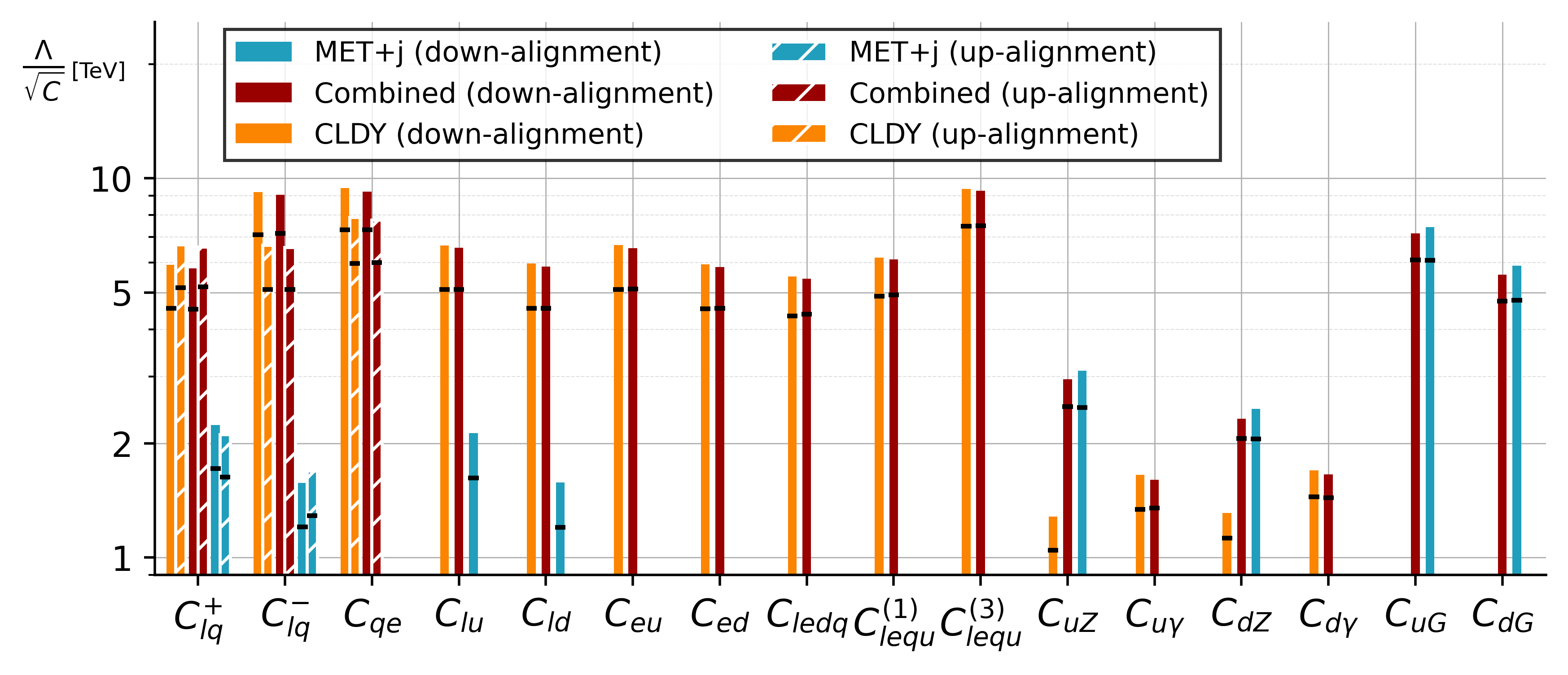}
  \caption{95\% limits on $\Lambda/\sqrt{C}$ in the lepton-flavor universal scenario for the quark indices $i,j=1,2$ from the combined fit (red) and the individual fits of the CLDY (orange) and MET+j (blue) processes. Limits in down-alignment (up-alignment) are shown as solid (dashed) lines.Black bars indicate the limits from single operator fits.}
  \label{fig:results_LU_12}
\end{figure}

We perform a fit for each quark flavor combination $i ,j, i\neq j$ individually. 
Credible intervals of the WCs can be translated into a bound on $\Lambda$, since the fits effectively only constrain $C/\Lambda^2$. 
We show these $95\%$ bounds on $\Lambda/\sqrt{C}$ from the lepton-flavor universal fit for the $i,j=1,2$ quark combination in Fig.~\ref{fig:results_LU_12}. The bounds in the down-alignment are shown as solid lines, while the bounds in the up-alignment are shown as dashed lines. Since the choice of basis does not affect most coefficients, we only show the bounds in the down-alignment basis for those\footnote{The differences arising from rotations of the dipole coefficients and $C_{ledq}$ fall within the uncertainties of this analysis. For $C_{lequ}^{(1)}$ and $C_{lequ}^{(3)}$, rotations result in differences of a few percent between the up- and down-alignment for the $i,j=1,2$ fit, owing to significant $uu$ contributions. For the $i,j=1,3$ and $i,j=2,3$ cases, the differences remain below the uncertainty threshold.}. 
The 95\% CL limits from single operator fits are indicated by the black bars.
We list the 95\% credible intervals in the LU scenario for completeness in Tab.~\ref{tab:CL_LU}.

Note that for the dipole coefficients as well as the scalar and tensor coefficients, we include $C_{ij}$ and $C_{ji}$ as independent degrees of freedom since the corresponding operators are not hermitian. The bounds are, however, identical because the operators contribute as ${|C_{ij}|^2 + |C_{ji}|^2}$ to the cross section, so that we only show one bound for each pair of coefficients. In total, the fit includes 25 free parameters for the $i,j=1,2$ quark-flavor combination: the sixteen operators from
Tab.~\ref{tab:operators}, nine of which -- all dipoles, one tensor and two scalars -- count double.

We see from Fig.~\ref{fig:results_LU_12} that the bounds on the four-fermion coefficients are dominated by CLDY data (orange), while the dipole coefficients, except for $C_{q\gamma}$, are stronger constrained by MET+jet (blue). This can be understood by the different kinematic regimes probed by the two processes. CLDY primarily probes the invariant mass spectrum ($m_{\ell \ell}$) of the dilepton pair, focusing on the high-energy tails that extend well beyond the $Z$-boson mass, $M_Z$. In this region, the SMEFT contributions of four-fermion operators are significantly enhanced by $\hat s$.
In contrast, MET+j is measured differentially in $\emiss$ which is dominated by the transverse momentum ($P_T$) distribution of the $Z$ boson. As the $\emiss$ spectrum effectively integrates over the invariant mass of the dineutrino pair, most events are clustered around $m_{\nu \bar \nu} \sim M_Z$, due to the resonance of the on-shell $Z$ bosons and the inverse correlation to the $P_T$. This on-shell enhancement increases the MET+j cross-section for the electroweak dipole operators, leading to stronger bounds compared to CLDY, 
where this enhancement is not present.

The bounds on the gluon dipole from MET+j are further strengthened by the fact that it predominately arises from gluon fusion. As illustrated in Fig.~\ref{fig:gluonDipole}, there is only one quark in the initial state, so that there is not necessarily a suppression due to the PDFs of the second and third generation quarks. In CLDY, in contrast, both quarks need to be present in the initial state so that the PDF suppression is more pronounced.

For the coefficients with two left-handed quark fields, $C_{lq}^\pm, C_{qe}$ there are visible differences between the limits in the up- and down-alignment. As expected from
 (\ref{eqn:Lagr_up}), (\ref{eqn:Lagr_down}), the bounds on the $C_{lq}^{+}$ coefficient are stronger in the up-alignment, since more contributions to the CLDY process arise due to 
 flavor mixing. By the analogous argument, the bounds on $C_{lq}^{-}$ and $C_{qe}$ are stronger in the down-alignment.

\begin{figure}[h]
 \centering
 \includegraphics[width=0.48\textwidth]{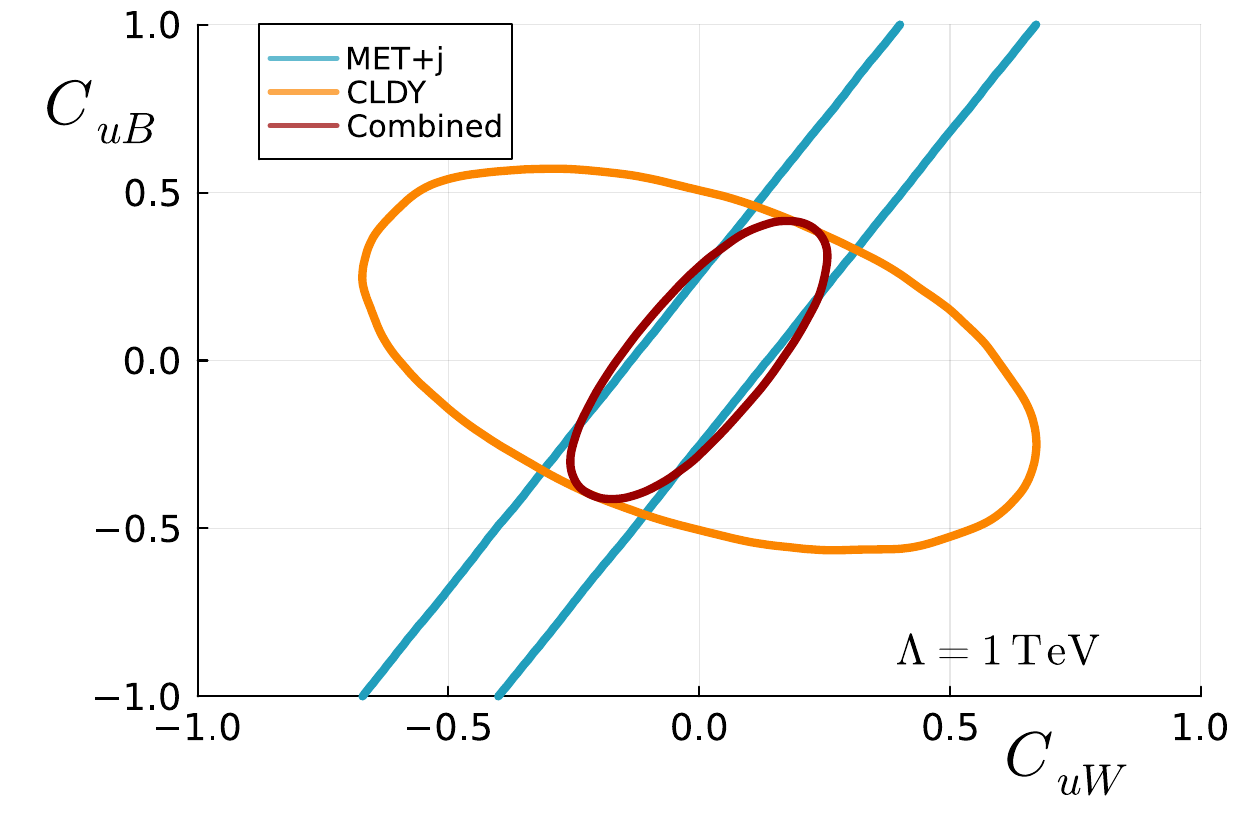}
 \includegraphics[width=0.48\textwidth]{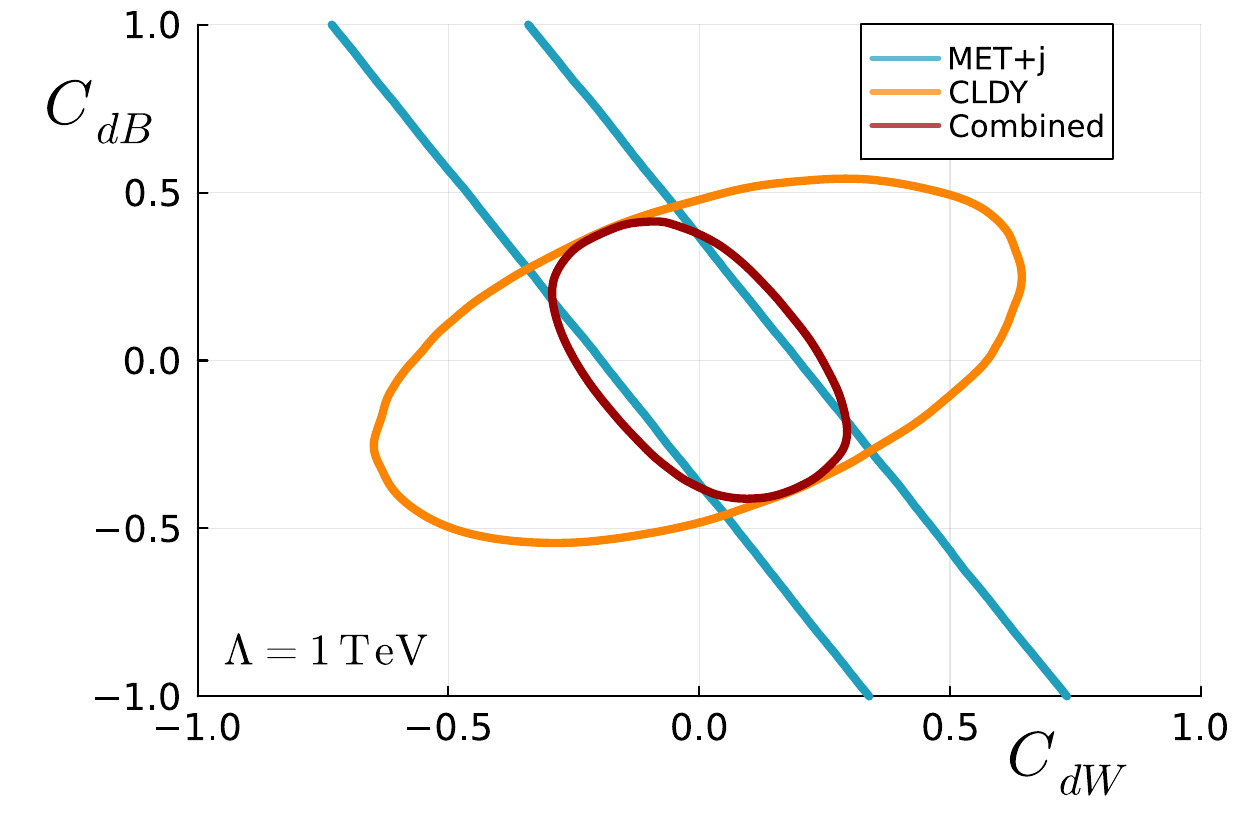}
 \caption{95\% credible contours for the dipole coefficients for the $i,j=1,2$ quark combination in the $C_{uB}-C_{uW}$ plane (left) and the $C_{dB}-C_{dW}$ plane (right). The fits are performed in the lepton-flavor universal scenario
 for $\Lambda= 1$ TeV.}
 \label{fig:results_dipoles_12}
\end{figure}

We highlight the synergies between CLDY and MET+j for the electroweak dipole operators. The 95\% credible contours in the $C_{uW}$-$C_{uB}$ plane (left panel) and
$C_{dW}$-$C_{dB}$ plane (right panel) are shown in Fig.~\ref{fig:results_dipoles_12} for the $i,j=1,2$ quark combinations. 
The combined analysis (red) significantly reduces the allowed parameter space, demonstrating complementarity between the CLDY (orange) and MET+j (blue) processes. For both 
up- and down-quark sectors, the addition of MET+j data shrinks the uncertainty regions, particularly along the $C_{qZ}$ direction.

Fig.~\ref{fig:results_LU_12} suggests rather counterintuitively that the reach of the combined fit is slightly worse than the one from the individual fits. This is likely an artefact of the Bayesian fitting, where the combined fit can be more conservative than the individual fits due to marginalization effects.

We also observe that the single operator fits give weaker bounds than the global analyses. 
This is an effect of the marginalization process in the global fit, where the 95\% credible intervals are obtained by marginalizing over all other coefficients. 
The posterior distributions typically  concentrate around one center, and drop towards the edges. On the other hand, single operator fits  yield rather double-peak-type 
posteriors. Their resulting 95\% probability interval is  wider than the one from the global analyses, and thus gives weaker bounds.
To understand how this happens consider  a simplified  toy system with two coefficients $C_{a,b}$ and two observables, $obs_1=|C_a|^2+|C_b|^2$ and $obs_2=C_b$.
In a fit to $C_a$ only, with $C_b=0$, the magnitude of $C_a$ is constrained, leaving  a twofold  ambiguity as the  sign  is not resolved. On the other hand, in a global analysis 
the interplay of the coefficients under the constraints from $obs_{1,2}$ together allows to resolve the ambiguity and give a posterior distribution that  has one peak,
rather than two.
We verified this behaviour explicitly by comparing selected posterior distributions from single and global fits.

\subsection{ Operators involving $b$-quarks \label{sec:b}}

\begin{figure}[hb]
  \centering
  \includegraphics[width=0.85\textwidth]{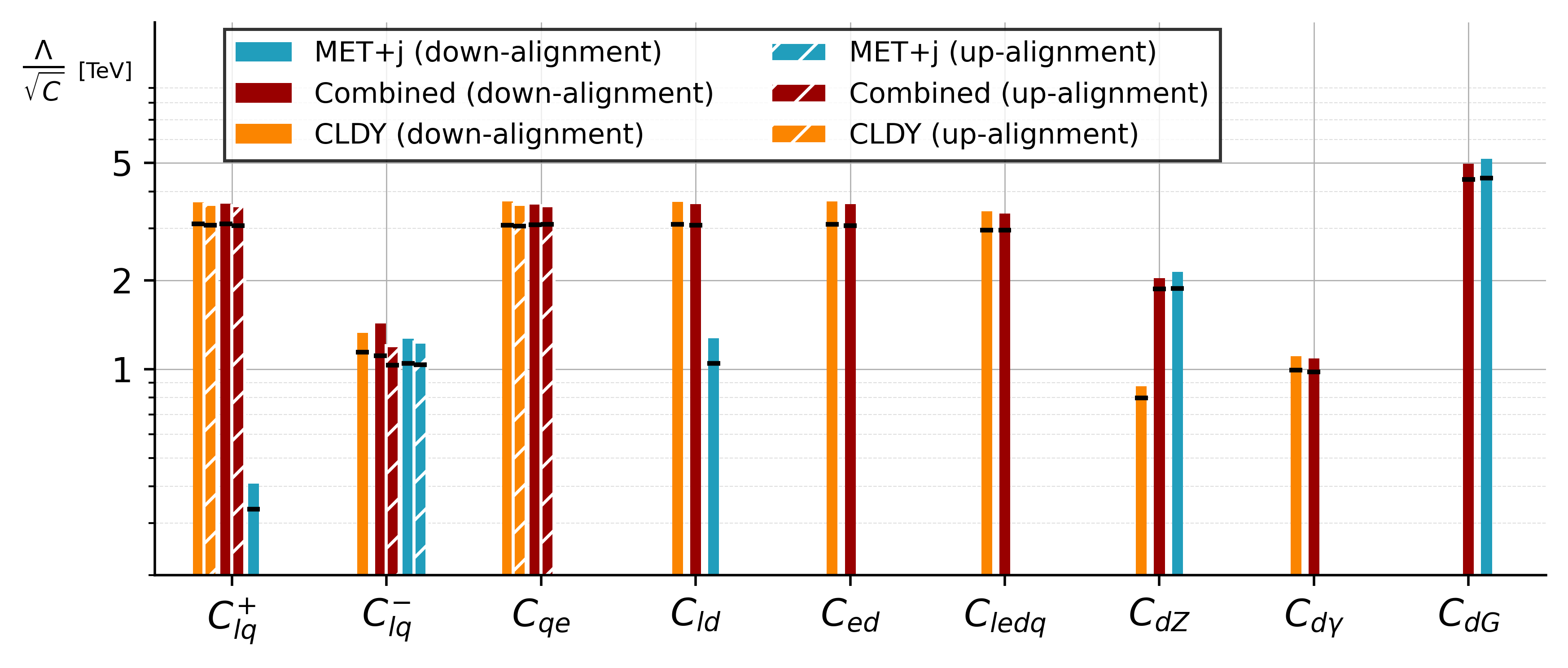}
  \caption{95\% limits on $\Lambda/\sqrt{C}$ in the lepton-flavor universal scenario for {$i,j=1,3$},
   see Fig.~\ref{fig:results_LU_12}.}
  \label{fig:results_LU_13}
\end{figure}

Since the top-quark as the $SU(2)_L$-partner of the $b$-quark is not accessible due to the PDF suppression, the fits involving the third generation differ from the $i,j=1,2$ one.
Firstly, as they are not probed, we do not include operators with right-handed up-type quarks in the fits of the $13$ and $23$ quark combinations, reducing the number of free parameters to 13. 
Secondly, since the $i,j=1,3$ and $i,j=2,3$ quark combinations only induce down-type quark contributions in DY, there are no bounds on $C_{lq}^{-}$ from CLDY and no bounds on $C_{lq}^{+}$ from MET+j in the up-alignment, see (\ref{eqn:Lagr_up}). In the down-alignment, see (\ref{eqn:Lagr_down}), on the other hand, the rotations induce 
other flavor combinations. 
Specifically, $C^{+}_{lq}$ induces up-type quark contributions to MET+j whereas $C^{-}_{lq}$ induces up-type quark contributions to CLDY.
Considering PDF and CKM hierarchies the dominant effect in $i,j=1,3$ stems from initial $uu$ and $uc$ contributions, which are of a similar magnitude. The $uc$ initial state contributes at order $\lambda^2$ in the Wolfenstein parameter $\lambda \simeq 0.2$, while the $uu$ initial state is of order $\lambda^3$. The latter is compensated because the $u$ quark PDF is enhanced compared to $c$ quark initial states and, as a flavor-diagonal term, further features interference terms with the SM. For $i,j=2,3$, the leading effect is from initial $uc$ as well as $cc$, contributing at order $\lambda^3$ and $\lambda^2$, respectively. The $cc$ initial state is PDF suppressed compared to the $uc$ initial state, but the CKM suppression is weaker and it interferes with the SM.

\begin{figure}[h]
 \centering
 \includegraphics[width=0.48\textwidth]{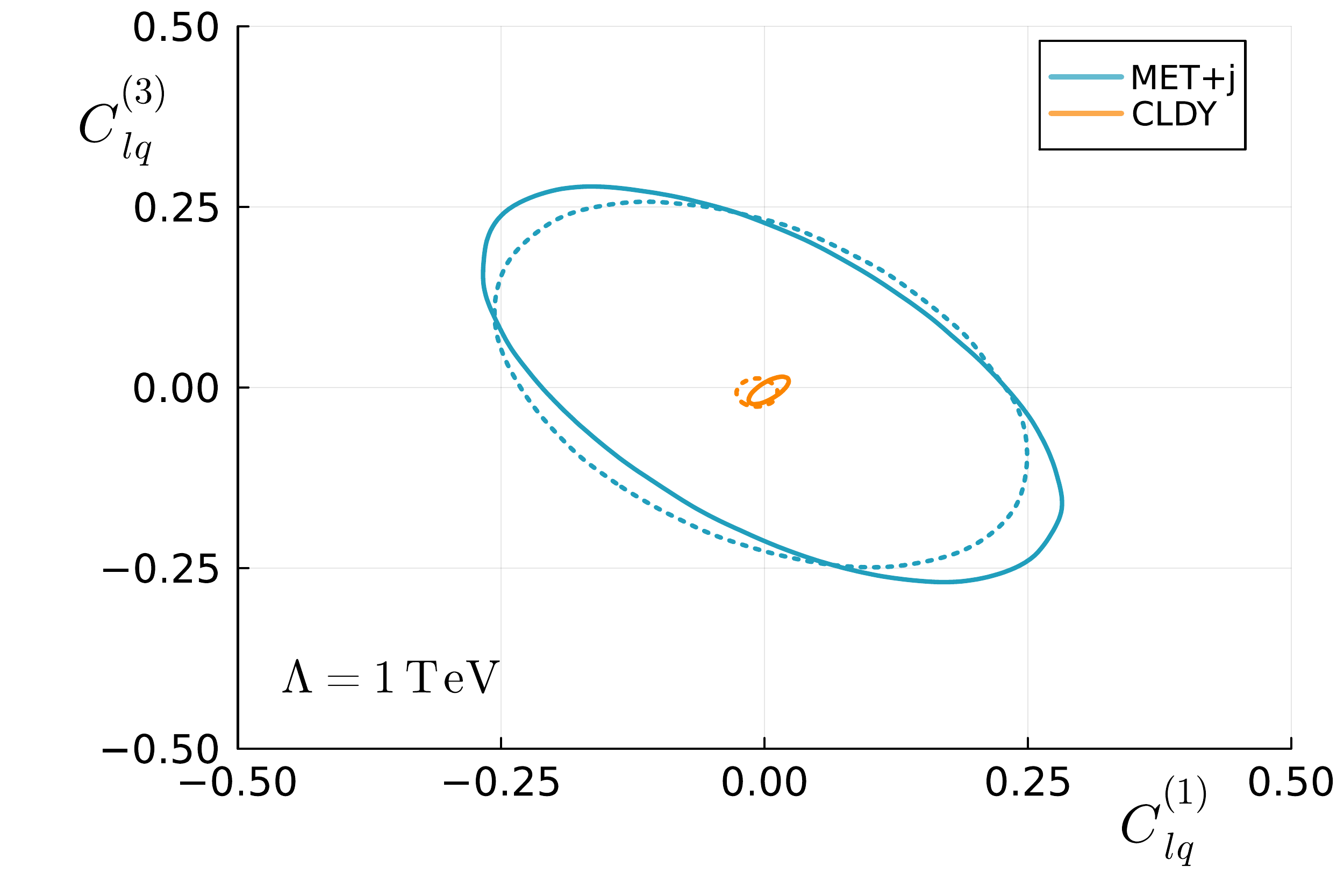}
 \includegraphics[width=0.48\textwidth]{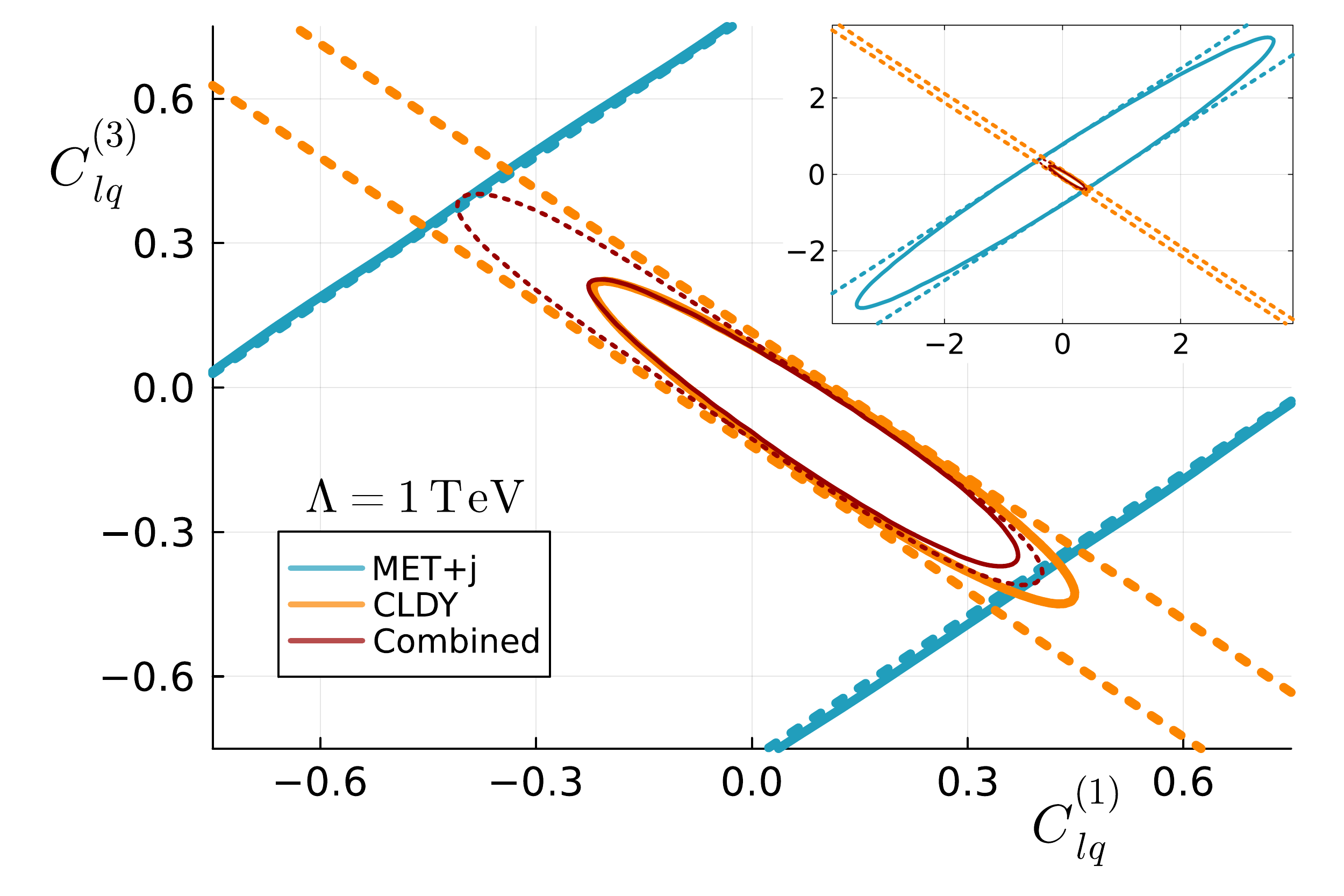}
 \caption{95\% credible contours for the $C_{lq}^{(1)}$ and $C_{lq}^{(3)}$ coefficients for the $i,j=1,2$ (left) and $i,j=1,3$ (right) quark combinations. The fits are performed in the lepton-flavor universal scenario for $\Lambda= 1$ TeV. Limits in down-alignment (up-alignment) are shown as solid (dashed) lines.}
 \label{fig:results_lq}
\end{figure}

\begin{figure}[h]
  \centering
  \includegraphics[width=0.48\textwidth]{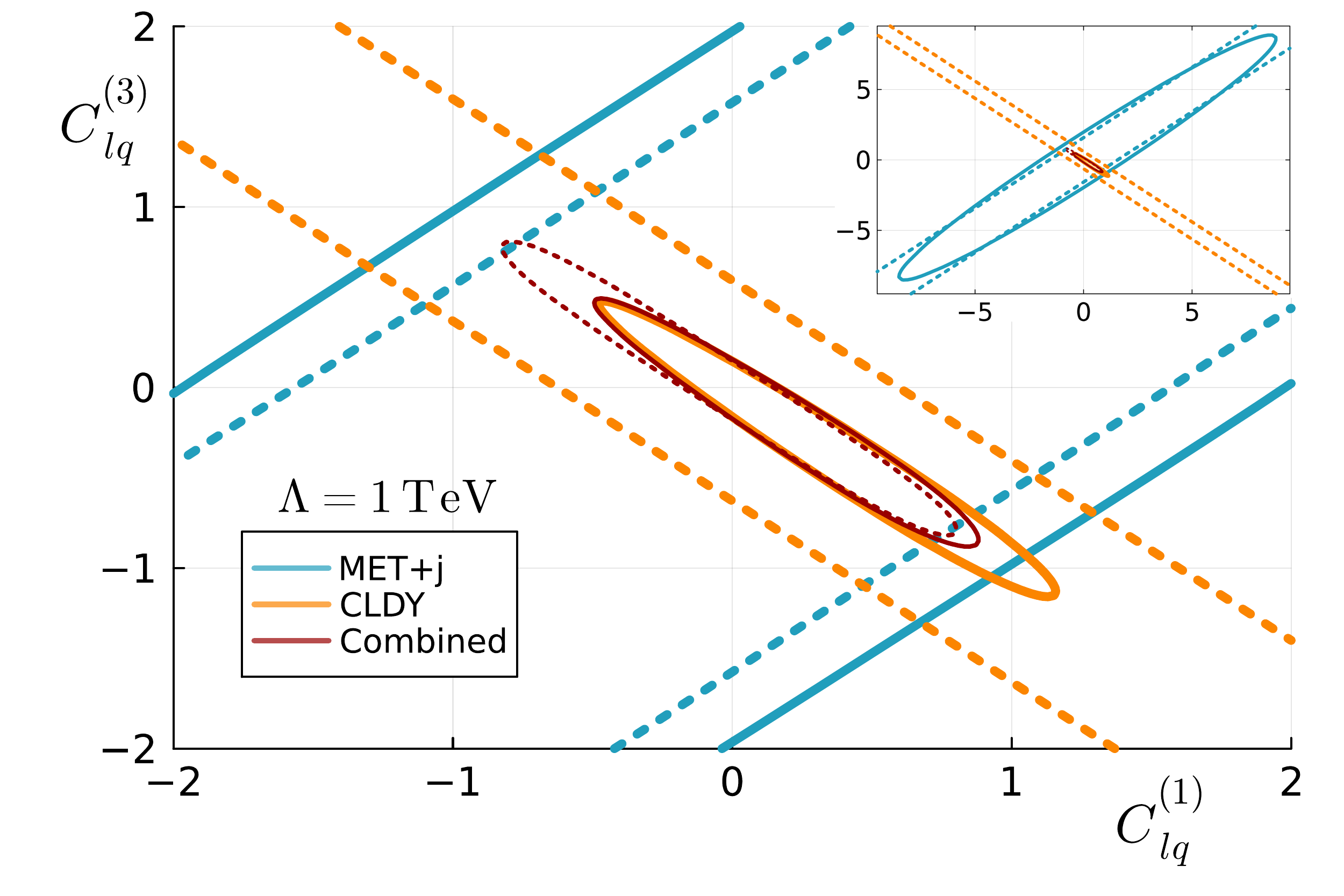}
  \caption{95\% credible contours for the $C_{lq}^{(1)}$ and $C_{lq}^{(3)}$ coefficients for the $i,j=2,3$ quark indices. The fits are performed in the LU scenario assuming $\Lambda= 1$ TeV. The results in the down (up) alignment are shown as solid (dashed) lines, see Fig.~\ref{fig:results_lq}.}
  \label{fig:results_lq_23}
\end{figure}

We go on and present results for $i,j=1,3$.
The fit results for $i,j=2,3$ are given in the appendix~\ref{app:auxiliary_limits} in Figs.~\ref{fig:results_LU_23},\ref{fig:results_lq_23}. They are qualitatively similar to the $i,j=1,3$ case, and differ mainly due to the different PDFs and CKM elements.
For the $i,j=1,3$ quark indices, the results for LU lepton flavor are shown in Fig.~\ref{fig:results_LU_13}.
Besides the improved constraints in the dipole sector from the joint analysis already discussed for $i,j=1,2$,
we find that flat directions in four-fermion couplings that arise in the individual fits, can be resolved.
To illustrate this we show in Fig.~\ref{fig:results_lq} the 95\% credible contours of $C_{lq}^{(1)}$ and $C_{lq}^{(3)}$ for the $i,j=1,3$ (right panel) and for comparison for 
the $i,j=1,2$ (left panel) quark combination in the down-alignment (solid lines) and up-alignment (dashed lines). Not shown for $i,j=1,2$ is the combined fit,
as it essentially overlaps with the bounds from CLDY (orange), which dominates the fit.
While for $i,j = 1,2$ both $C_{lq}^{(1)}$ and $C_{lq}^{(3)}$ coefficients can be constrained by the individual processes, there are flat directions in $C_{lq}^{(1)}$ and $C_{lq}^{(3)}$ for 
operators with $b$-quarks in the up-alignment. The combined fit (red) is able to resolve these.

The 95\% credible contours of $C_{dW}$ and $C_{dB}$ for the $i,j=1,3$ and $i,j=2,3$ quark indices are shown in Fig.~\ref{fig:results_dipoles_app} in the appendix~\ref{app:auxiliary_limits}. The operators
$O_{\underset{i3}{uW}}$, $O_{\underset{i3}{uB}}$ and $O_{\underset{i3}{uG}}$ are not constrained, as the top-quark is not accessible in leading order Drell-Yan.

\subsection{Comparison of lepton-flavor patterns}
\label{sec:LFV_vs_LU}

\begin{figure}[h]
  \centering
  \includegraphics[width=0.85\textwidth]{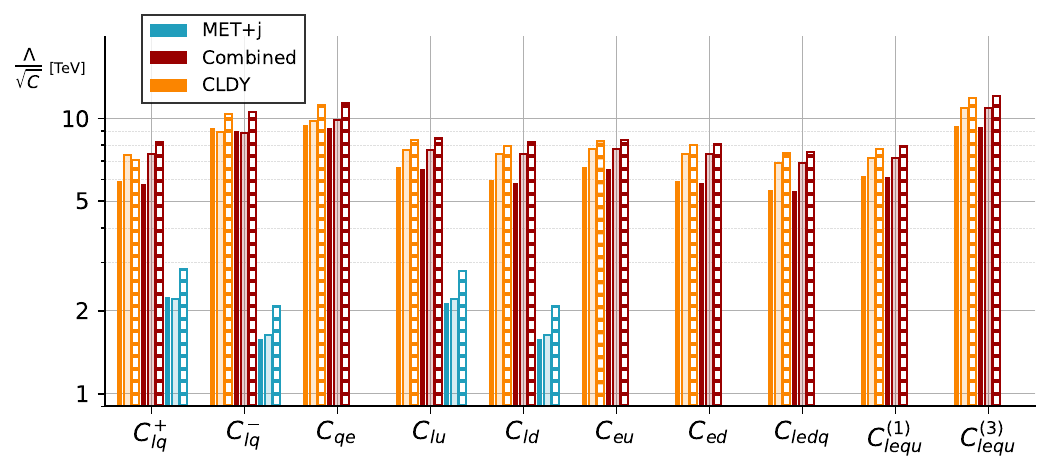}
  \caption{95\% limits on $\Lambda/\sqrt{C}$ in the lepton-flavor universal (solid), lepton-flavor violating (shaded) and democratic (striped) scenario for the quark indices $i,j=1,2$ from the combined fit (red) 
  and the individual fits of the CLDY (orange) and MET+j (blue) in down-alignment.}
  \label{fig:results_LFV_12}
\end{figure}

We also consider a LFV scenario, where the lepton-flavor off-diagonal elements are assumed to be equal, cf.~Sec.~\ref{sec:leptons}, as well as a democratic pattern where lepton-flavor violating and lepton-flavor conserving couplings are assumed to have the same strength. We compare the bounds on these LFV and democratic couplings to the LU ones in Fig.~\ref{fig:results_LFV_12}. Shown are the 95\% credible limits on $\Lambda/\sqrt{C}$ for the $i,j=1,2$ quark combination in the down-alignment for the combined and the individual fits. The results for the other quark indices are given in the appendix~\ref{app:auxiliary_limits}. 
Since the dipole operators do not generate LFV contributions, they are not included in this fit. We see that the bounds on the LFV coefficients (shaded) are slightly stronger than the LU ones (solid), which can be understood by the smaller background in experimental analyses of LFV. The bounds are, however, at a similar order of magnitude. 

The democratic bounds (striped) are stronger than the LU or LFV ones, as they are effectively a combination of the LU and LFV bounds. The relative increase is especially pronounced if the LU and LFV bounds are of similar order of magnitude.

\subsection{Lepton-flavor specific fits}
\label{sec:LF_specific}

\begin{figure}[h]
  \centering
  \includegraphics[width=0.85\textwidth]{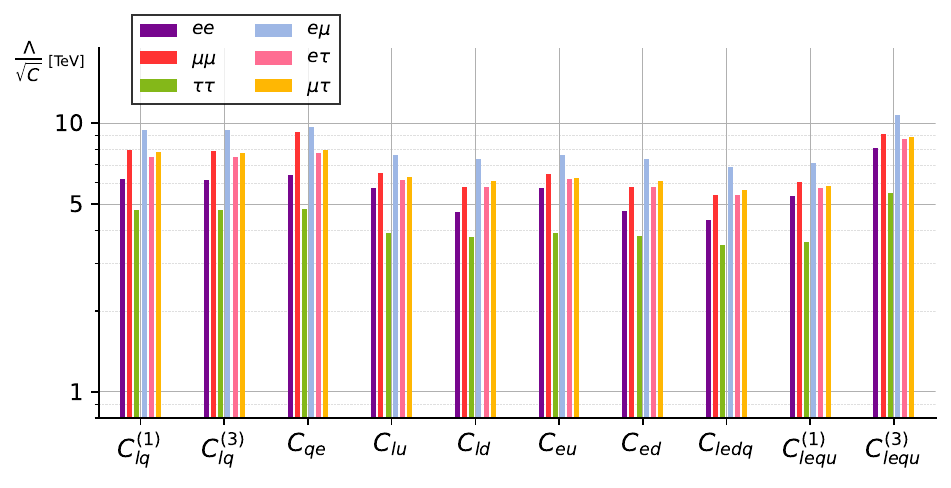}
  \caption{95\% limits on $\Lambda/\sqrt{C}$ for the lepton-flavor specific couplings. We show the results of the combined fit for the quark indices $i,j=1,2$ in the down-alignment.}
  \label{fig:results_LF_specific_12}
\end{figure}

\begin{figure}[h]
  \centering
  \includegraphics[width=0.8\textwidth]{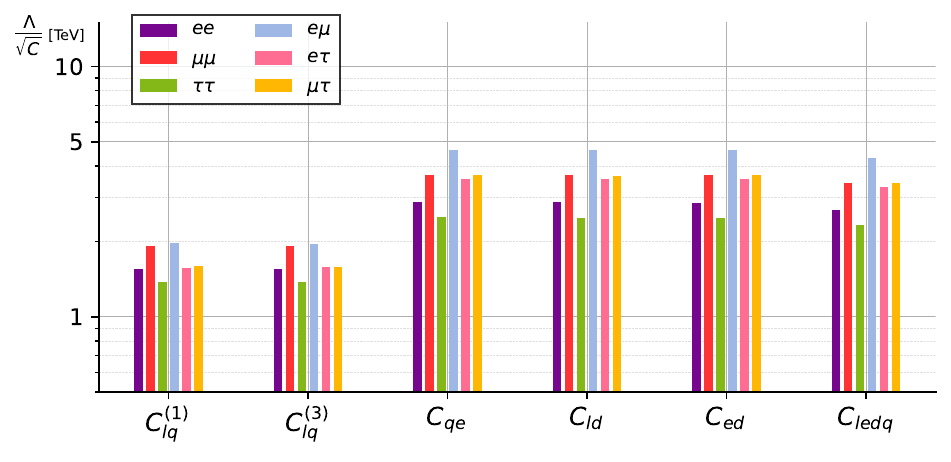}
  \caption{95\% limits on $\Lambda/\sqrt{C}$ for the lepton-flavor specific couplings. We show the results of the combined fit for the quark indices $i,j=1,3$ in the down-alignment, 
  see Fig.~\ref{fig:results_LF_specific_12}.}
  \label{fig:results_LF_specific_13}
\end{figure}

\begin{figure}[h]
  \centering
  \includegraphics[width=0.8\textwidth]{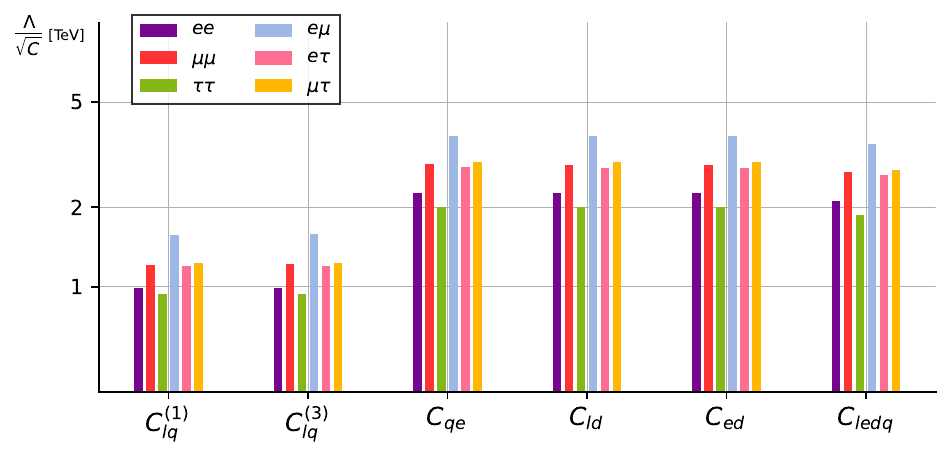}
  \caption{95\% limits on $\Lambda/\sqrt{C}$ for the lepton-flavor specific couplings. We show the results of the combined fit for the quark indices $i,j=2,3$ in the down-alignment, 
  see Fig.~\ref{fig:results_LF_specific_12}.}
  \label{fig:results_LF_specific_23}
\end{figure}

We also perform lepton-flavor specific fits, where each coefficient $ C_{\alpha \beta ij}$ is constrained individually. We show the 95\% credible limits on $\Lambda/\sqrt{C}$ for the $i,j=1,2$ quark combination in Fig.~\ref{fig:results_LF_specific_12} in down-alignment. The results for the other quark indices are shown in Figs.~\ref{fig:results_LF_specific_13},\ref{fig:results_LF_specific_23}.

We see that the strongest bounds arise for the $e\mu$ coefficients, since this channel provides an experimentally easily accessible signature and exhibits only a small SM background. The bounds on the $e\tau$ and $\mu\tau$ coefficients are weaker, as the $\tau$ lepton is more challenging to reconstruct.
The best flavor diagonal bounds are on $\mu\mu$ coefficients, as the muon can be reconstructed well at colliders and the experimental analysis includes bins up to very high energy scales, where the SMEFT contributions profit strongly from the energy enhancement. The bounds on the $ee$ coefficients are weaker, as the electron provides a more challenging experimental signature. The $\tau\tau$ coefficients are the least constrained, since the $\tau$ lepton is very challenging to reconstruct due to its rapid decay. The $\tau\tau$ channel moreover features a large SM background, which further decreases the sensitivity.

\subsection{Synopsis of DY fits}
\label{sec:synopsis}

The strongest bounds exist on the four-fermion tensor operator $O_{lequ}^{(3)}$, due to the energy enhancement and as it contributes with a large prefactor to the effective coefficients in Eq.~\eqref{eqn:C_4F_ll}. In the LU scenario for the quark indices $i,j=1,2$, the sensitivity of this operator on the new physics scale $\Lambda$ is $\sim 10\,$TeV for a $C\sim 1$ contribution. The bounds on the dipole operators are weaker, but the gluon dipole can still be constrained to $\Lambda \sim 8\,$TeV due to the full energy enhancement in the MET+j process. The bounds on the $C_{lq}^{(1/3)}$ and $C_{qe}$ are also strong, with $\Lambda \sim 9\,\text{TeV}$ in the LU scenario in the down-alignment, because both, the $uc$ and $ds$ process are induced simultaneously. For these operators, we observe a noticeable difference between the up- and down alignment bases.
The limits on the electroweak dipole operators are the weakest, between $\sim 1-3$ TeV. This is comparable to the limits for light quarks obtained in a related analysis of $pp \to Z +j$~\cite{Gauld:2024glt}. We like to stress that this process is also sensitive to gluon dipoles, 
which unlike the electroweak ones, enter in a fully energy-enhanced manner, as shown in App.~\ref{app:G_dipoles}.

In the $i,j=1,3$ or $i,j=2,3$ quark combinations, there are flat directions in the $C_{lq}^{(1/3)}$ in the up-alignment that are resolved by combining CLDY and MET+j. Here, the bounds on $C_{dG}$ are the stronger than the four-fermion operators due to the different PDF suppressions.

The LFV contributions are slightly more constrained than the lepton-flavor conserving LU contributions, because the experimental analysis feature less background and are thus more sensitive to the SMEFT coefficients. Because the democratic scenario combines the lepton-flavor universal and lepton-flavor violating contributions, it yields overall bounds that are
more stringent than those in the LU and LFV cases.
The lepton-flavor specific fits provide the strongest bounds on the $e\mu$ coefficients, while the $\tau\tau$ coefficients are the least constrained.

\section{The benefits of rare decays}
\label{sec:Flavor}

We combine the constraints on dipole operators obtained from the CLDY and MET+j processes with those from flavor observables. 
The corresponding analysis of semileptonic four-fermion operators has recently been presented in \cite{Hiller:2024vtr}.
The interest in the dipole operators is in their energy enhancements in Drell-Yan processes, as well as their sensitivity to chirality flipping interactions at loop-level.
A well-known UV-model that can induce large dipole operators  is the R-parity conserving MSSM, for instance. Here, the chiral flip can come from the
heavy super partners  of the gauge bosons, which together with sizable FCNC squark-mixing ("$A$-terms")  causes an enhancement for this type of operator and not for other, chirality conserving ones, or $\Delta F=2$ operators,  e.g~\cite{Giudice:2012qq}. Note, due to the second Higgs doublet one needs to be in the decoupling limit to be able to match the MSSM onto SMEFT~\cite{Dawson:2022cmu}.

We begin reviewing the current constraints on the dipole operators from flavor observables.
Flavor observables are usually analyzed in the weak effective theory (WET). The WET dipole operators are defined as 
\begin{equation}
  \begin{aligned}
    &O_{\underset{ij}{7}}^{(\prime)} = \frac{e}{16\pi^2} m_i \left( \bar q^i_{L(R)} \sigma^{\mu\nu} q^j_{R(L)}\right) F_{\mu\nu} \,, \\
    &O_{\underset{ij}{8}}^{(\prime)} = \frac{g_s}{16\pi^2} m_i \left( \bar q^i_{L(R)} \sigma^{\mu\nu} T^A q^j_{R(L)}\right) G^A_{\mu\nu} \,,
  \end{aligned}
  \label{eqn:WET_dipoles}
\end{equation}
with the electric charge $e$, the strong coupling constant $g_s$ and quark masses $m_i$. The corresponding Lagrangian reads
\begin{equation}
  \mathcal{L}_{\text{WET}} = \frac{4G_F}{\sqrt{2}} \sum_{O^{ij}} \lambda_{\text{CKM}}^{ij} C^{ij} O^{ij} \,,
  \label{eqn:WET_Lagrangian}
\end{equation}
with the CKM factor $\lambda_{\text{CKM}}^{ij} = V_{ti} V_{tj}^*$ for the down-type quarks to indicate the flavor suppression of the SM contributions,
 and $\lambda_{\text{CKM}}^{ij}=1$ for $cu$ operators. For the latter the GIM-mechanism implies vanishing  dipole coefficients  at the weak scale 
 and only very small effective coefficients at the charm mass scale induced by charged-current four-quark operators \cite{deBoer:2017que}. Therefore, as customary in new physics analyses in charm, we do not  put any non-trivial  CKM pre-factors to rescale FCNC NP-coefficients.

Due to significant hadronic uncertainties, precise, systematic limits on gluon dipole couplings for the first and second generation quarks are not available.
We get an estimate for the upper limit by demanding that the contribution from the BSM couplings $C_{8}^{(\prime)}$ to hadronic decays does not exceed the SM contribution as follows:
The dipole operators induce a contribution to, say, $D \to K^+ K^-$, $D \to K^{*+} K^-$, or $K \to \pi \pi$, $K \to \pi \pi \pi$, at tree-level, with color-suppression.
The reason for considering not just two pseudoscalars in the final state is that their decay amplitude probes $C_{8}-C_{8}^{\prime}$, and we want to avoid large cancellations.
The hadronic decays are induced in the SM at tree-level by the weak interaction, color-favored, but singly Cabibbo-suppressed. 
We require the ratio of the dipole to SM amplitude, 
\begin{equation}
r \sim C_{8}^{(\prime)} \frac{ \alpha_s}{4 \pi} \frac{\lambda_{\text{CKM}}^{ij}}{N_c \lambda} \lesssim 1
\end{equation}
to be less than one. Here, $N_c=3$ is the number of colors.
The corresponding limit is of order ${\mathcal{O}(10)}$ for $cu$ transitions and ${\mathcal{O}(10^4)}$ for $ds$ transition, where the difference stems from the CKM factors.
We are aware that this procedure is naive and subject to large hadronic uncertainties. In absence of other information it turns out to be useful to have an order of magnitude limit.
For the coefficient $C_{\underset{ds}{7}}$, an upper bound can be estimated by requiring that the new physics contribution to the decays $K_{S,L} \to \gamma \gamma$ amounts to less than 10\% of the SM amplitude, $\lvert C_{7}^{(\prime)} \rvert \lesssim 500$~\cite{Mertens:2011ts}.
Limits on the new physics contributions from FCNC decays are summarized in Tab.~\ref{tab:flavor_constraints}. The coefficients are given at the scale $\mu=m_c$ for $uc$, $\mu=m_b$ for $db$ and $sb$ and at 1 GeV for $ds$. 

\begin{table}[htb]
  \setlength{\tabcolsep}{15pt}
  \renewcommand{\arraystretch}{1.2}
  \centering
  \begin{tabular}{c | c c c c}
    & $C_{7}$ & $C_{7}^{\prime}$ & $C_{8}$ & $C_{8}^{\prime}$ \\ 
    \toprule
    $uc$ & [-0.26, 0.18]~\cite{Gisbert:2024kob} & [-0.18, 0.25]~\cite{Gisbert:2024kob} & $\lesssim {\mathcal{O}}(10)$ & $\lesssim {\mathcal{O}}(10)$ \\
    $ds$ & $\lesssim 500$ ~\cite{Mertens:2011ts} & $\lesssim 500$ ~\cite{Mertens:2011ts} & $\lesssim {\mathcal{O}}(10^4)$ & $\lesssim {\mathcal{O}}(10^4)$ \\
    $db$ & [-0.07, 0.11]~\cite{Bause:2022rrs} & [-0.18, 0.16]~\cite{Bause:2022rrs} & [-0.88, 1.44]~\cite{Bause:2022rrs} & [-1.16, 1.13]~\cite{Bause:2022rrs} \\
    $sb$ & [-0.02, 0.01]~\cite{Alguero:2021anc} & [-0.01, 0.02]~\cite{Alguero:2021anc} & [-1.20, -0.40]~\cite{Mahmoudi:2023upg} & [-1.60, 1.00]~\cite{Mahmoudi:2023upg} \\
  \end{tabular} 
  \caption{Limits on FCNC WET dipole coefficients from flavor data. Shown are 1$\sigma$ confidence intervals from global analyses and order of magnitude upper limits from the requirement that new physics does not exceed the SM contribution, see text.}
  \label{tab:flavor_constraints}
\end{table}

To connect the bounds from flavor and DY observables, we match the SMEFT onto the WET at the scale $\mu=m_Z$ at tree-level
\begin{equation}
  \begin{aligned}
    C_{7}^{( \prime)} &= \frac{8\pi^2v^3}{\sqrt{2\pi \alpha_{em}}m_i \Lambda^2 \lambda_{\text{CKM}}^{ij}} \left( \cos{\theta_W} C_{\underset{ij(ji)}{dB}} - \sin{\theta_W} C_{\underset{ij(ji)}{dW}} \right) \,, \\
    C_{8}^{( \prime)} &= \frac{8\pi^2v^3}{\sqrt{2\pi \alpha_{s}}m_i \Lambda^2 \lambda_{\text{CKM}}^{ij}} C_{\underset{ij(ji)}{uG}} \,.
  \end{aligned}
\end{equation}

For the Renormalization Group Equation (RGE) running, we employ the one-loop RGE for the SMEFT coefficients~\cite{Jenkins:2013zja,Jenkins:2013wua,Alonso:2013hga} as well as the WET coefficients~\cite{Aebischer:2017gaw}. The numerical integration is performed using the Python package \texttt{Wilson}~\cite{Aebischer:2018bkb}.
$C_{7}^{(\prime)}$ and $C_{8}^{(\prime)}$ mix under WET RGE, such that constraints on $C_{7}^{(\prime)}$ at a low scale constrain $C_{7}^{(\prime)}$ and $C_{8}^{(\prime)}$ at higher scales. Therefore, constraints on $C_{7}^{(\prime)}$ from flavor observables can be employed to constrain the SMEFT coefficients $C_{\underset{ij}{qW/B}}$ as well as $C_{\underset{ij}{qG}}$ at the high scale $\Lambda$.

\begin{figure}[h]
  \centering
  \includegraphics[width=0.48\textwidth]{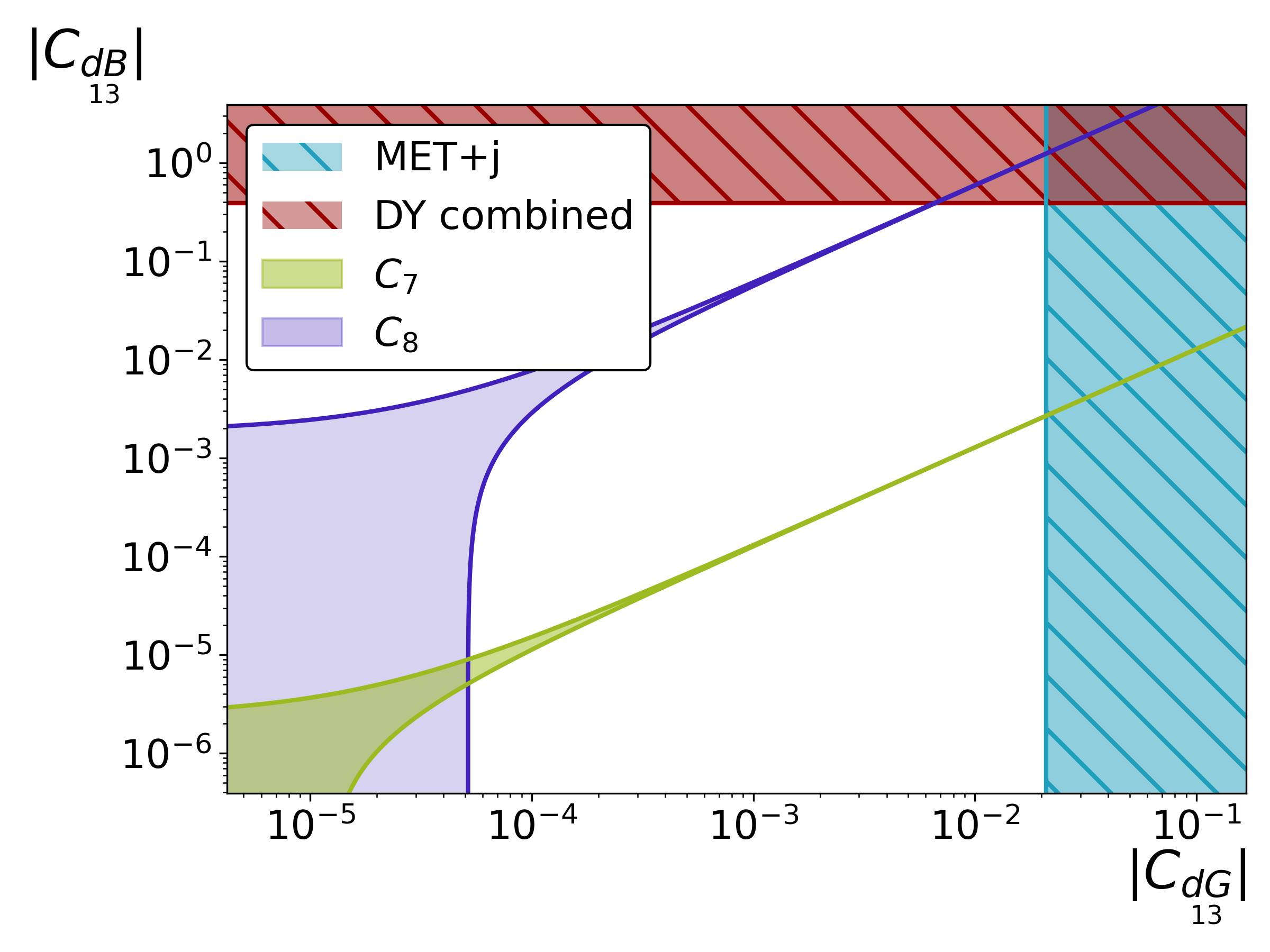}
  \includegraphics[width=0.48\textwidth]{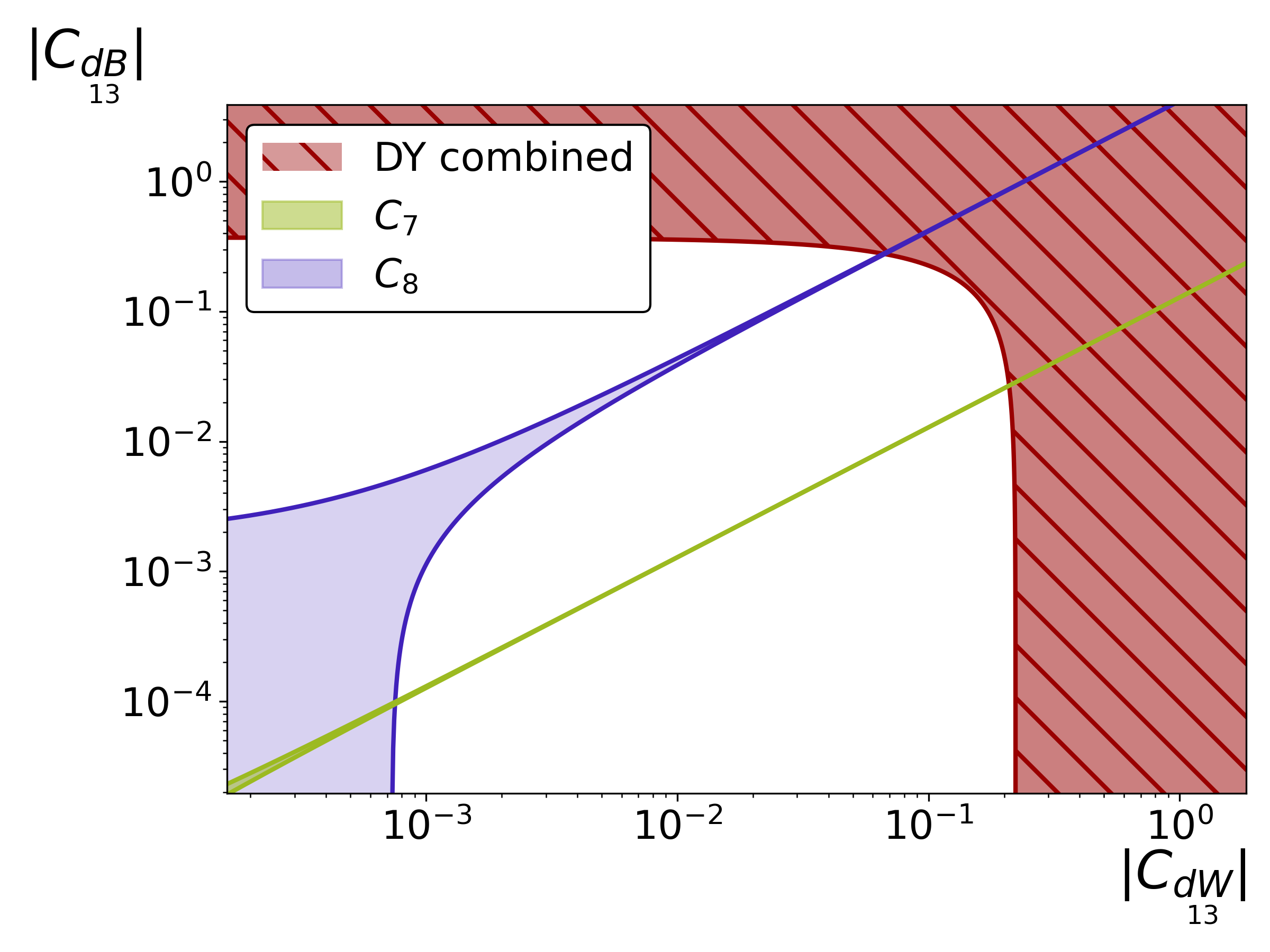}
  \caption{Flavor and collider constraints on the dipole coefficients in the 
  $C_{dG}$-$C_{dB}$ plane (left) for fixed ${C_{dW}=0}$, and in the $C_{dW}$-$C_{dB}$ plane (right) with ${C_{dG}=0}$ for $i,j=1,3$ quark FCNCs.
  The red (blue) regions are excluded by CLDY and MET+j data (MET+j data) at 68\% CL, while the green and purple regions represent the 1$\sigma$ regions allowed by constraints on $C_7$ and $C_8$, respectively.}
  \label{fig:comparison_flavor}
\end{figure}

While the SMEFT dipole operators span a three dimensional parameter space, the flavor constraints only provide two linearly independent constraints.
In Fig.~\ref{fig:comparison_flavor}, we illustrate the interplay between flavor and collider constraints on the dipole coefficients for a $db$ transition by showing slices of the parameter space where we set one of the three dipole coefficients to zero, ${C_{dW}=0}$ (left panel) and ${C_{dG}=0}$ (right panel).
While this demonstrates the principle how better bounds can be obtained using complementary information, the combined bounds from flavor and DY cannot be inferred from these slices and a full 3D analysis is required. Results of the latter are presented in Tab.~\ref{tab:combined_constraints}. When bounds improve compared to the DY only fit, we indicate the improvement in brackets as the ratio $C_{\text{combined}}/C_{\text{DY}}$ in percent.

\begin{table}[h]
  \renewcommand{\arraystretch}{1.2}
  \centering
  \begin{tabular}{l | l l l l}
    & $uc$ & $ds$ & $db$ & $sb$ \\ 
    \toprule
    $C_{qB}$ & [0.063, 0.063] (27\%) & [-0.095, 0.095] (41\%) & [-0.13, 0.13] (25\%) & [-0.24, 0.24] (37\%) \\
    $C_{qW}$ & [-0.039, 0.039] (30\%) & [-0.058, 0.058] (39\%) & [-0.080, 0.080] (28\%) & [-0.15, 0.15] (38\%) \\
    $C_{qG}$ & [-0.010, 0.010] & [-0.0054, 0.0054] (34\%) & [-0.0078, 0.0078] (37\%) & [-0.015, 0.015] (28\%) \\
  \end{tabular} 
  \caption{$1\sigma$ intervals of the combined constraints on the SMEFT coefficients from DY and flavor data for $\Lambda = 1\,\text{TeV}$. For the coefficients that improve compared to the DY only fit, we indicate the improvement in brackets as the ratio $C_{\text{combined}}/C_{\text{DY}}$ in percent.}
  \label{tab:combined_constraints}
\end{table}

As can be seen, the combination of Drell Yan with rare decay constraints is beneficial. Bounds are improved in most cases by a factor of $\sim 3$. For the $uc$ transition, there is no improvement on $C_{uG}$ because the flavor constraints, in particular the one on $C_{8}$, are relatively weak~\footnote{For CP-violating couplings the limits in charm are stronger by 3-4 orders of magnitude.}.

There is clear gain from global analysis: the drawback of the rare hadron decays is that they have significant hadronic uncertainites, especially in the gluon dipole contributions. Moreover, they directly probe from the electroweak sector only the photon.
On the other hand, these low energy contraints significantly  improve on the DY-only limits, see Table IV. In addition, there is recent debate about a new physics contamination in the extraction of pdfs from high energies, that would impact the high-$p_T$ data.

\section{Future collider sensitivities}
\label{sec:FutureCollider}

We turn to the prospects of future hadron colliders.
To estimate the reach we employ a simplified extrapolation of our bounds based on the approximate statistical significance. For this, we focus on a single inclusive bin, where we discuss in particular the effects of the choice of its lower edge $m_{\text{cut}}$ on the sensitivity.

The specifications of the future collider setups considered in this analysis are summarized in Tab.~\ref{tab:FutureCollider} together with the current LHC data on Drell Yan~\cite{CMS:2021ctt,ATLAS:2020zms}. In particular, we investigate the prospects of the High-Luminosity LHC (HL-LHC)~\cite{ZurbanoFernandez:2020cco}, the High-Energy LHC (HE-LHC)~\cite{FCC:2018bvk}, and the Future Circular Collider (FCC-hh)~\cite{Benedikt:2022kan}. Similar studies have been performed recently on monophoton and displaced vertex signatures for the $e^+ e^-$ Future Circular Collider~\cite{Bolton:2025tqw}.
\begin{table}[htb]
  \renewcommand{\arraystretch}{1.2}
  \setlength{\tabcolsep}{15pt}
  \centering
  \begin{tabular}{c c c }
    Collider & $\sqrt{s} \, / \,\text{TeV} $ &$\mathcal{L} \, / \, \text{ab}^{-1} $ \\ \toprule 
    LHC & $13$ & $0.14$ \\
    HL-LHC & $14$ & $3$ \\
    HE-LHC & $27$ & $15$ \\ 
    FCC-hh & $100$ & $20$ \\
  \end{tabular} 
  \caption{Center-of-mass energy and integrated luminosity of the LHC and future hadron colliders considered in this analysis.}
  \label{tab:FutureCollider}
\end{table}

We focus on four-fermion operators and the gluon dipole operator given in Tab.\ref{tab:operators}, along with the observables $m_{\mu \mu}$ and $\emiss$ presented in Tab.\ref{tab:data_sets}, since these provide the most significant constraints in our analysis.
To estimate a future reach on the WCs, we limit our analysis to a single inclusive high invariant mass or high-$p_T$ bin. This assumption is motivated by the fact that both types of operators are fully energy-enhanced, so that the the constraints will be dominated by the high energy tails. The statistical significance $Z$ is calculated following Ref.~\cite{Cowan:2010js} as
\begin{equation}
    Z = \sqrt{2 \left( \left(N_{\text{sig}} +N_{\text{bgd}} \right) \log \left(1 +\frac{N_{\text{sig}}}{N_{\text{bgd}}} \right) - N_{\text{sig}} \right)}
    \approx \frac{N_{\text{sig}} }{ \sqrt{N_{\text{bgd}}}} \,,
  \label{eqn:Stat_significance}
\end{equation}
where $N_{\text{sig}}$ and $N_{\text{bgd}}$ denote the number of signal and background events, respectively. The approximation holds under the assumption that $N_{\text{sig}} \ll N_{\text{bgd}}$, which is reasonable since we assume that possible NP signals are small due to the suppression by the high scale $\Lambda^{-4}$.

The significance in general depends on the lower edge of the inclusive bin. We consider this as a free parameter, which we refer to as $m_{\text{cut}}$. To maximise the sensitivity, we investigate the dependence of the significance on this parameter and choose the value of $m_{\text{cut}}$ that maximizes it. As an example, we discuss the operator $C_{\underset{22 12}{lu}}$ for the process $pp \to \mu^- \mu^+$ in the following. This observable is measured depending on invariant mass of the muon pair $m_{\mu \mu}$, so that $m_{\text{cut}}$ corresponds to $m_{\mu \mu}^{\text{min}}$ for the bin in this case.

Neglecting acceptance and efficiency effects of the detector, the number of background events can be written as
\begin{equation}
  \begin{aligned}
  N_{\text{bgd}} &= \mathcal{L} \sum_{i,j}\int_{\tau_{\text{cut}}}^1 \frac{\mathrm{d} \tau}{\tau} L_{ij}(\tau,\mu_F^2) \, \hat \sigma^{\text{SM}}_{ij}\left( m_{\text{cut}}^2 , \tau s\right) \\
  &= \frac{\mathcal{L}}{s} A^{\text{SM}} \int_{\tau_{\text{cut}}}^1 \frac{\mathrm{d} \tau}{\tau^2 } L_{\text{SM}}(\tau,\mu_F^2) \,,
  \end{aligned}
  \label{eqn:Nbackground}
\end{equation}
where $\tau_{\text{cut}}= m_{\text{cut}}^2/s$, $\mathcal{L}$ denotes the integrated luminosity and $L_{ij}(\tau,\mu_F^2) $ are the parton luminosity functions defined in Eq.~\eqref{eqn:PLFs}.
To factor out its leading energy dependence, we parameterized the SM parton cross section, following Eq.~\eqref{eqn:DY_SM_xsec} with $\hat s = \tau s$, as
\begin{equation}
  \hat \sigma_{ij}^{\text{SM}} = A^{\text{SM}} \alpha_{ij} \frac{1}{\tau s} \,,
\end{equation}
and defined
\begin{equation}
  L_{\text{SM}}(\tau,\mu_F^2) = \sum_{i,j}\alpha_{ij} L_{ij}(\tau,\mu_F) \,,
\end{equation}
with $ A^{\text{SM}} = \frac{4 \pi \alpha^2}{9}$ and $\alpha_{ij} = \frac{\hat \sigma_{ij}^{\text{SM}} \tau s}{A^{\text{SM}}}$.
Similarly, the number of signal events can be written as 
\begin{equation}
  N_{\text{sig}} = \mathcal{L} A^{\text{BSM}} s \frac{\lvert C_{\underset{22 12}{lu}} \rvert^2}{\Lambda^4} \int_{\tau_{\text{cut}}}^1 \mathrm{d}\tau L_{uc}(\tau,\mu_F^2) \,,
  \label{eqn:Nsignal}
\end{equation}
with $A^{\text{BSM}} = \frac{2}{144 \pi} $, with a factor $2$ due to combinatorics of the initial states.
The dependence in Eq.~\eqref{eqn:Nsignal} on $\tau$ and $s$ differs compared to the background events~\eqref{eqn:Nbackground} due to the energy enhancement of the SMEFT operators, as can be seen in Eq.~\eqref{eqn:DY_4F_xsec}. 
Setting the factorization scale to $\mu_F^2 = \tau s$, the significance for a $C_{uc}$ SMEFT coefficient can be written as 
\begin{equation}
  Z = \tilde A \sqrt{\mathcal{L}} \, \lvert C_{\underset{22 12}{lu}} \rvert^2 \, \frac{ s^{3/2} }{\Lambda^4} \frac{\int_{\tau_{\text{cut}}}^1 \mathrm{d} \tau L_{uc}(\tau,\tau s) }{ \sqrt{\int_{\tau_{\text{cut}}}^1 \frac{\mathrm{d} \tau}{\tau^2 } L_{\text{SM}}(\tau,\tau s) }} \,,
  \label{eqn:significance}
\end{equation}
where $\tilde A$ is an overall normalization factor. From Eq.~\eqref{eqn:significance}, it is evident that the significance $Z$ scales with the square root of the integrated luminosity $\mathcal{L}$ and the energy to the power of~$3/2$. The significance will thus increase with increasing center-of-mass energy and integrated luminosity.

\begin{figure}
  \centering 
  \includegraphics[width = 0.8\textwidth]{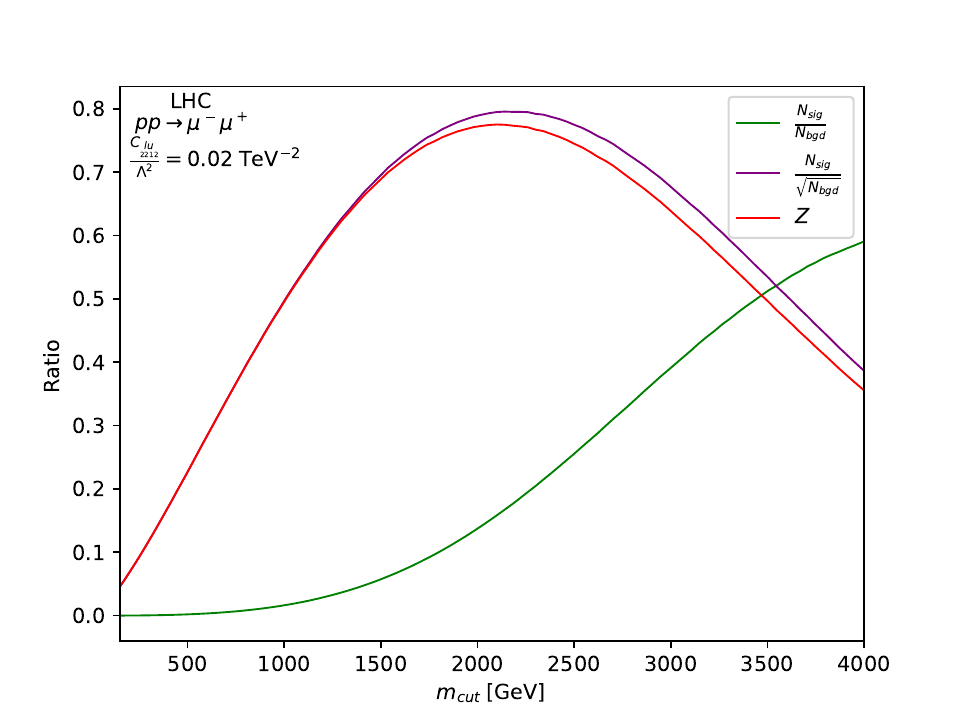}
  \caption{The ratios $\frac{N_{\text{sig}}}{\sqrt{N_{\text{bgd}}}} $, $\frac{N_{\text{sig}}}{N_{\text{bgd}}} $ as well as the significance $Z$ defined in (\ref{eqn:Stat_significance}) as a function of the inclusive cut $m_{\text{cut}}$.
  The ratio $\frac{N_{\text{sig}} }{ N_{\text{bgd}}} $ (green) grows monotonously with the cutoff value $m_{\text{cut}}$, while the other two peak around a preferred intermediate value.}
  \label{fig:RatioPlot}
\end{figure}

While the ratio $\frac{N_{\text{sig}} }{ N_{\text{bgd}}}$ increases steadily with increasing values for $m_{\text{cut}}$, this is not the case for $\frac{N_{\text{sig}} }{\sqrt{N_{\text{bgd}}}}$. 
As $N_{\text{bgd}}$ decreases, $Z$ initially grows. However, the significance will eventually decrease again, as the background statistics become too small. This results in a peak in $Z$ at an intermediate value of $m_{\text{cut}}$, which depends on the relative factor of the integrated parton luminosity functions.
To illustrate this, we show the different ratios for the operator $O_{\underset{22 12}{lu}}$ in Fig.~\ref{fig:RatioPlot} for a value of $C_{\underset{22 12}{lu}}/\Lambda = 0.02\,\text{TeV}^{-2}$.

To determine the expected reach of future hadron colliders for a given $m_{\text{cut}}$, we calculate the significance $Z$ using Eq.~\eqref{eqn:significance} for the respective collider setup. We vary the parameter $C / \Lambda^2$ until $Z$ reaches $2\sigma$, corresponding to a 95\% confidence level bound. This calculation is performed using \mg, which evaluates the full expression in Eq.~\eqref{eqn:Stat_significance} while applying the basic selection cuts employed in Refs.~\cite{CMS:2021ctt,ATLAS:2021kxv} for the $m_{\mu \mu}$ and $\emiss$ observables, respectively.
Here, we use the MC-variant of the NNPDF4.0~PDF~sets~\cite{NNPDF:2021njg,Cruz-Martinez:2024cbz} as it leads to more stable results compared to the baseline set.

For the $p p \to \mu^+ \mu^-$ process, we consider the operator $O_{\underset{22 ij}{lu}}$($i\neq j$) as our signal contribution, while for the $\emiss$ observable of the MET+j process we consider $O_{\underset{ ij}{uG}}$($i \neq j$). 
For the latter, a similar significance to Eq.~\eqref{eqn:significance} can be defined, where the relation between the scaling variable and $E_T^{\text{miss},\text{cut}}$ is however more involved and additional initial states with gluons contribute.

We further assume that the background events $N_{\text{bgd}}$ of the MET+j analysis only arise from an intermediate $Z$-boson, whereas in the experimental analysis~\cite{ATLAS:2021kxv} additional backgrounds are considered.
The dominant background, through an intermediate $Z$-boson, accounts for about $70 \%$ in the highest bin of the ATLAS analysis \cite{ATLAS:2021kxv}, whereas diboson and vector boson fusion each acount for about $6\%$.
The largest contending background is given by W boson production of a charged lepton, without tagging the lepton, which accounts for about $15 \%$.
Estimating this background at a future collider is difficult at a future collider, as it depends on the efficiencies of the corresponding detector.
However all these processes are SM-like and the cross section falls for sufficiently high center-of-mass energies, similiar to Eq.~\eqref{eqn:Nbackground}.
Bearing in mind that we aim for a rough estimate of the reach of future colliders, only considering the primary background provides for a sufficient approximation.

\begin{figure}[htb]
  \centering 
  \includegraphics[width = 0.48\textwidth]{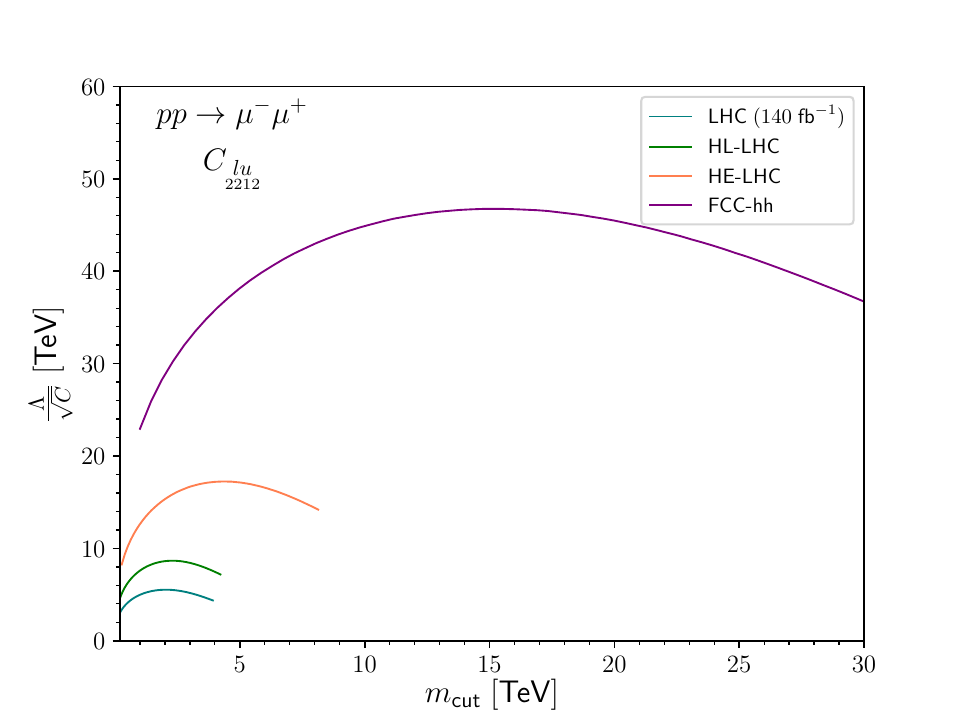}
  \includegraphics[width = 0.48\textwidth]{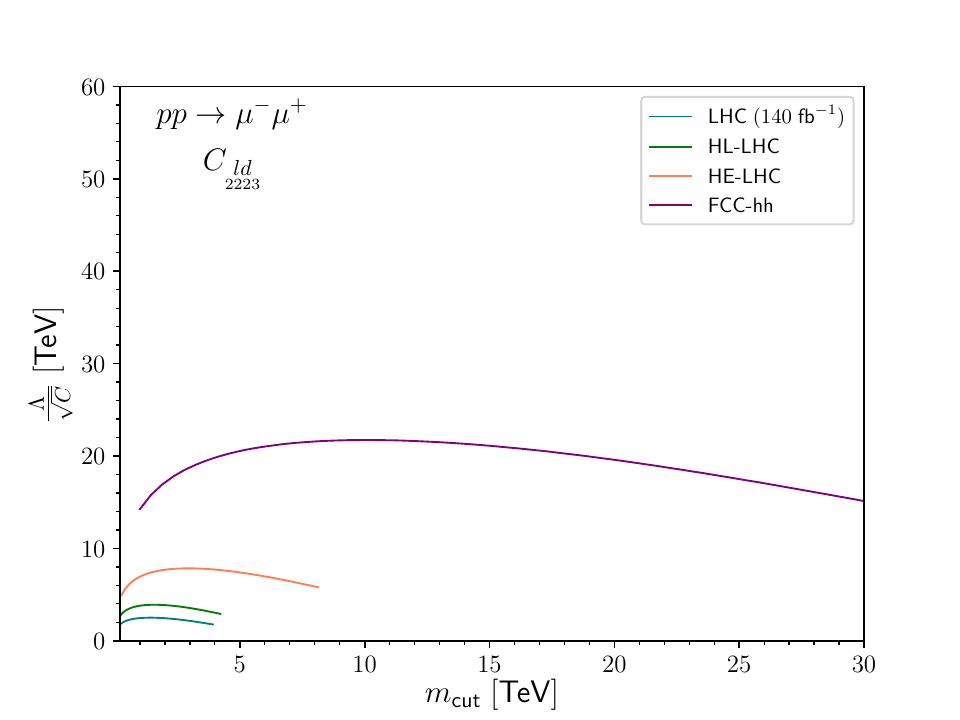}
  \includegraphics[width = 0.48\textwidth]{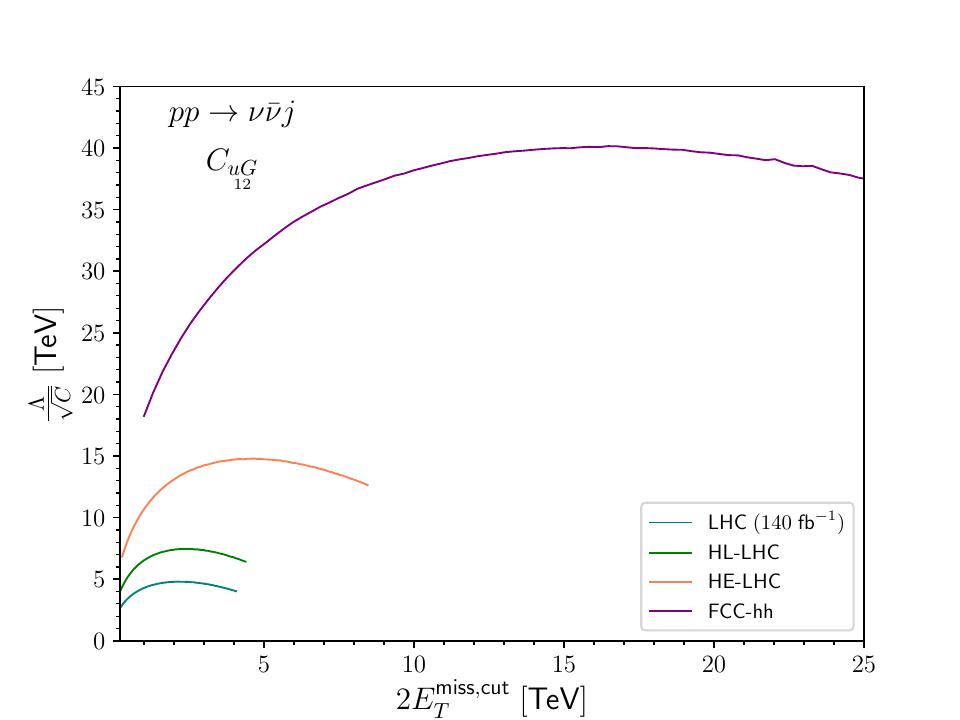}
  \includegraphics[width = 0.48\textwidth]{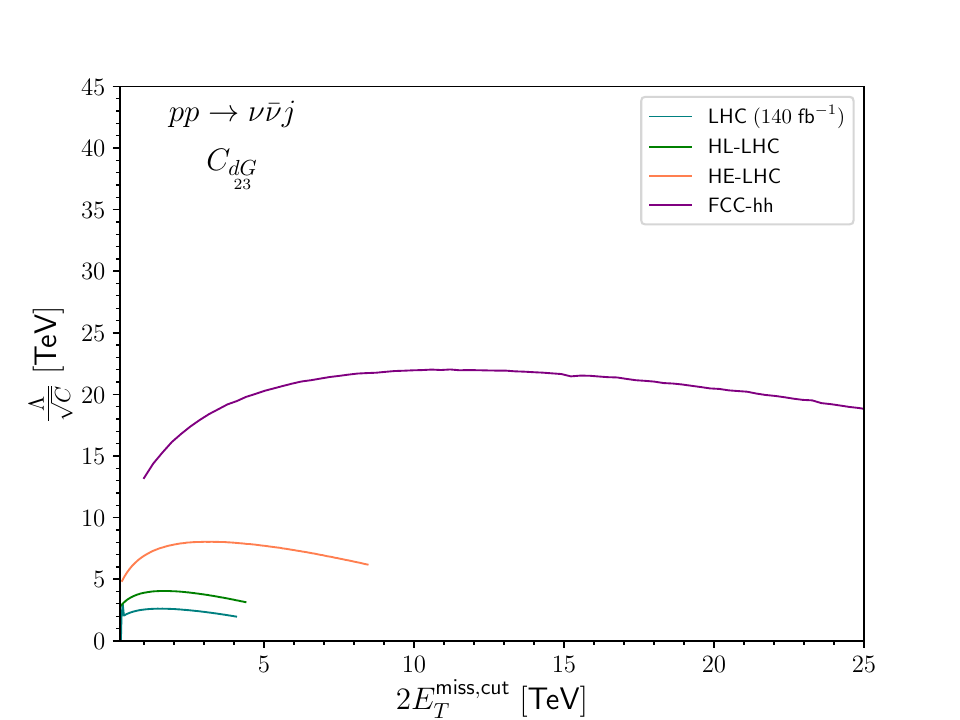}
  \caption{The estimated reach on the NP scale~$\Lambda$ as a function of the kinematic cut of the inclusive bin for the $m_{\mu \mu}$ observable (upper row) and the $\emiss$ observable (lower low) for a $uc$ transition (left column) and a $sb$ transition (right column). The considered signal processes are the $O_{\underset{22ij}{lu/d}}$ and the $O_{\underset{ij}{u/dG}}$ for $i \neq j$ contributions, respectively.}
  \label{fig:NPreach_uc_bs}
\end{figure}

\begin{figure}[h]
  \centering 
  \includegraphics[width = 0.48\textwidth]{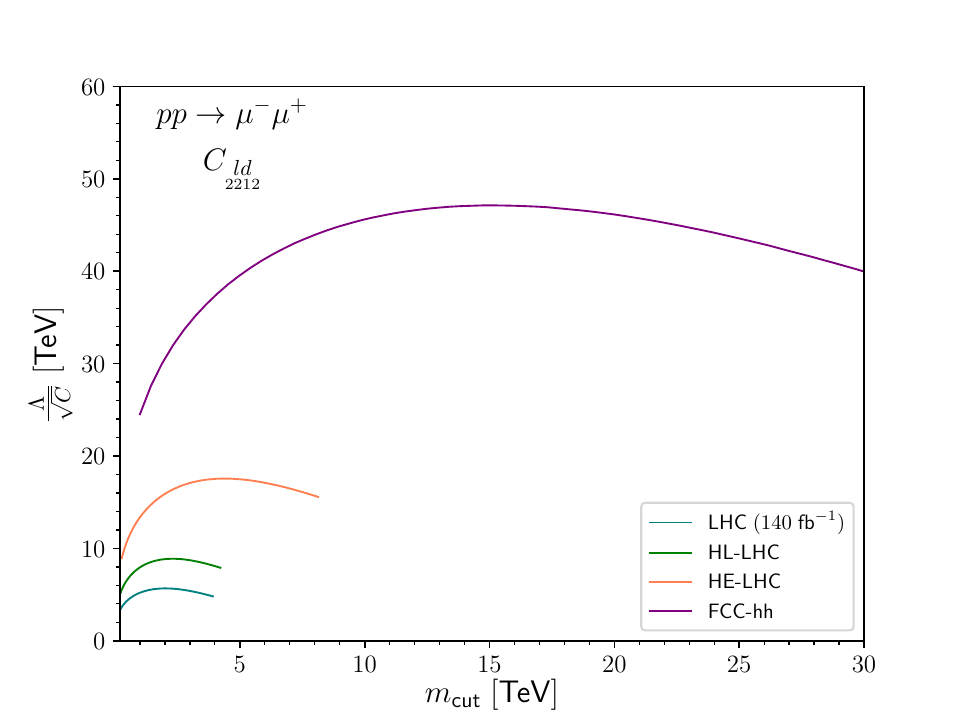}
  \includegraphics[width = 0.48\textwidth]{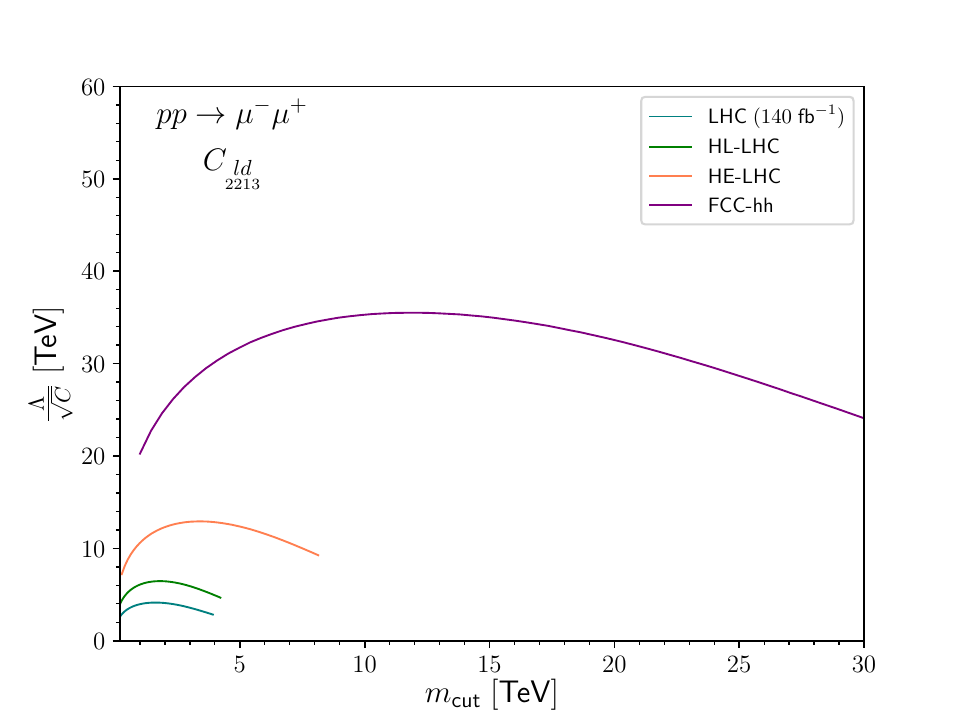}
  \includegraphics[width = 0.48\textwidth]{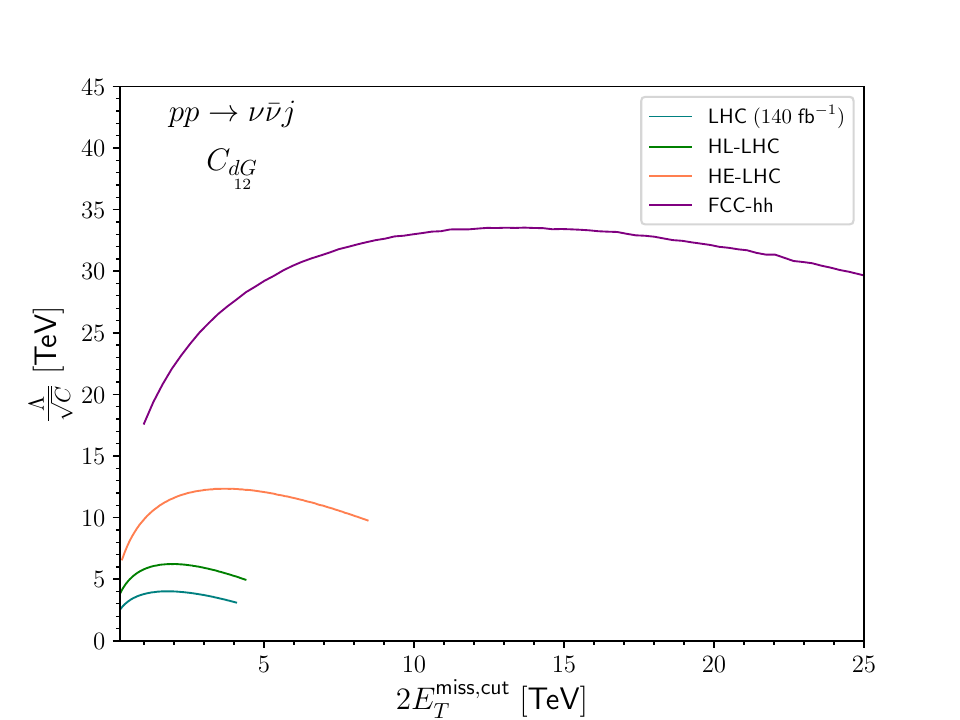}
  \includegraphics[width = 0.48\textwidth]{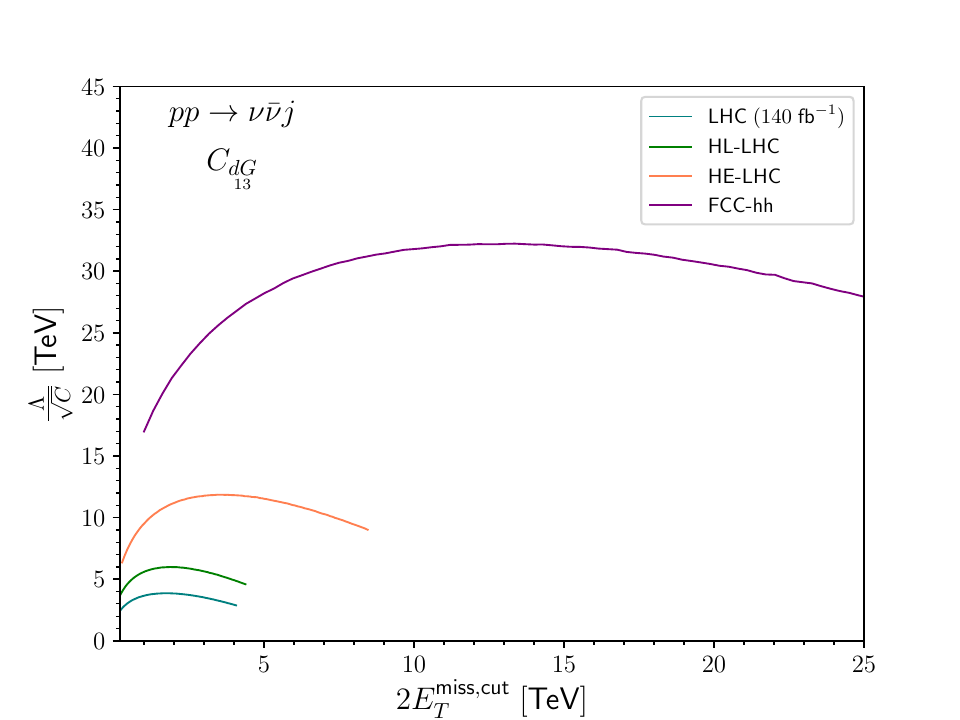}
  \caption{The estimated reach on the NP scale~$\Lambda$ as a function of the kinematic cut of the inclusive bin for the $m_{\mu \mu}$ observable (upper row) and the $\emiss$ observable (lower row) for a $ds$ transition (left column) and a $db$ transition (right column). The considered signal processes are the $O_{\underset{22ij}{ld}}$ and the $O_{\underset{ij}{dG}}$ contribution, respectively, see also Fig.~\ref{fig:NPreach_uc_bs}.}
  \label{fig:NPreach_ds_db}
\end{figure}

In Fig.~\ref{fig:NPreach_uc_bs}, we present the estimated 
95\% C.L. sensitivity on the $uc$ and $sb$ coefficients to $\Lambda$ as a function of the lower edge of the highest bin for the colliders listed in Tab.~\ref{tab:FutureCollider}. The results show that the statistical significance $Z$ peaks at an intermediate cutoff value, which corresponds to an optimal binning choice for this basic single-bin approach. This value differs significantly for the different quark transitions. The corresponding plots for the $ds$ and $db$ transitions are shown in Fig.~\ref{fig:NPreach_ds_db}.

\begin{table}[htb]
  \centering
  \resizebox{\columnwidth}{!}{
  \begin{tabular}{ |c| c c |c c c |c c c |c c c| c c c|} 
    \hline 
     & \multicolumn{2}{c|}{$C_{\underset{22 12}{lu}} $ } & &\multicolumn{2}{c|}{$C_{\underset{22 12}{ld}} $}  & &\multicolumn{2}{c|}{$C_{\underset{22 13}{ld}} $} & & \multicolumn{2}{c|}{$C_{\underset{22 23}{ld}} $}  \\ 
     \hline 
     & $\Lambda\; / \; \si{\TeV}$ & $m_{\mu \mu}^{\text{cut}} \; / \; \si{\TeV} $ &  & $\Lambda\; / \; \si{\TeV}$ & $m_{\mu \mu}^{\text{cut}} \; / \; \si{\TeV} $& & $\Lambda\; / \; \si{\TeV}$ & $m_{\mu \mu}^{\text{cut}} \; / \; \si{\TeV} $  & &  $\Lambda\; / \; \si{\TeV}$ & $m_{\mu \mu}^{\text{cut}} \; / \; \si{\TeV} $  \\
  LHC     & 5.5 &2.1    && 5.7   &2.0   && 4.1 & 1.6  && 2.5 &1.5 \\ 
  HL-LHC  & 8.7 &2.4    && 8.9   &2.4   && 6.5 & 1.8  && 3.9 &1.6 \\
  HE-LHC  & 17  &4.3    && 18    &4.6   && 13  & 3.4  && 7.8 &3.0 \\ 
  FCC-hh  & 47  &15   && 47    &15  && 36  & 12 && 22  &10 \\ 
  \hline
  & \multicolumn{2}{c|}{$C_{\underset{ 12}{uG}}$ } & &\multicolumn{2}{c|}{$C_{\underset{ 12}{dG}}$ } & &\multicolumn{2}{c|}{$C_{\underset{ 13}{dG}}$ } & & \multicolumn{2}{c|}{$C_{\underset{ 23}{dG}}$ } \\ 
  \hline
  & $\Lambda\; / \; \si{\TeV}$ & $E_T^{\text{miss},\text{cut}}\; / \; \si{\TeV} $ &  & $\Lambda\; / \; \si{\TeV}$ & $E_T^{\text{miss},\text{cut}}\; / \; \si{\TeV} $ & & $\Lambda\; / \; \si{\TeV}$ & $E_T^{\text{miss},\text{cut}}\; / \; \si{\TeV} $  & &  $\Lambda\; / \; \si{\TeV}$ & $E_T^{\text{miss},\text{cut}}\; / \; \si{\TeV} $ \\
  LHC     &  4.8    & 1.1   &&   4.0  & 0.90     && 3.9   &  0.89    && 2.6     & 0.81 \\ 
  HL-LHC  &  7.5    & 1.2   &&   6.2  & 1.0      && 6.0   &  0.9     && 4.0     & 0.78 \\
  HE-LHC  &  15     & 2.3   &&   12   & 1.8      && 12    &  1.7     && 8.0     & 1.6 \\ 
  FCC-hh  &  40     & 8.2   &&   34   & 6.8      && 32    &  6.6     && 22      & 5.6 \\ 
  \hline
  \end{tabular}
  }
\caption{Optimal cuts and the corresponding expected sensitivity on the NP scale~$\Lambda$ based on Eq.~\eqref{eqn:Stat_significance} for the coefficients $C_{\underset{22 ij}{lu/d}}$ and $C_{\underset{ij}{qG}}$ at present and future hadron colliders.}
\label{tab:FutureNP}
\end{table}
The optimal values for the cuts on the kinematic variables and the corresponding values for the sensitivity on $\Lambda$ are summarized in Tab.~\ref{tab:FutureNP}.
The results for the LHC are in good agreement with CLDY ones shown in Fig.~\ref{fig:results_LU_13}, but they undershoot the MET+j results. This can be understood for the $\emiss$ observable, due to the simplifications employed in this analysis.
Hence, the results for the MET+j process in Tab.~\ref{tab:FutureNP} should be seen as a relative improvement upon the LHC results, rather than an absolute reach.

In particular, the sensitivity of the FCC is about a factor of $8$ larger than at the LHC, while the HL-LHC and HE-LHC are expected to improve the bounds by factors of roughly $1.5$ and $3$, respectively. 
Furthermore, we observe that bounds on the $ds$ transitions are slightly better than the ones on the $uc$ transitions. This can be traced back to the usage of the MC variant of NNPDF4.0~PDF~sets~ in the computations. In this variant, the charm PDF is smaller compared to the baseline set, as it has already been observed in Ref.~\cite{Cruz-Martinez:2024cbz}.

\section{Conclusions}
\label{sec:conclusion}

We present a SMEFT-analysis of the total Drell-Yan process at the LHC, combining charged and neutral lepton production, $pp \to \ell^+ \ell^{(\prime)-}$ and $pp \to \nu \bar \nu j$. We show, as anticipated, that 
the combination strengthens the fit as it resolves flat directions, and allows to probe for a larger set of Wilson coefficients, cf. Figs.~\ref{fig:results_LU_12} and \ref{fig:results_dipoles_12}. Synergies are even stronger for fits with $b$-quarks, see Fig.~\ref{fig:results_lq}, since their $SU(2)_L$-partner, the top quark is not probed at leading order Drell-Yan.

\begin{figure}[h]
  \centering
  \includegraphics[width=0.8\textwidth]{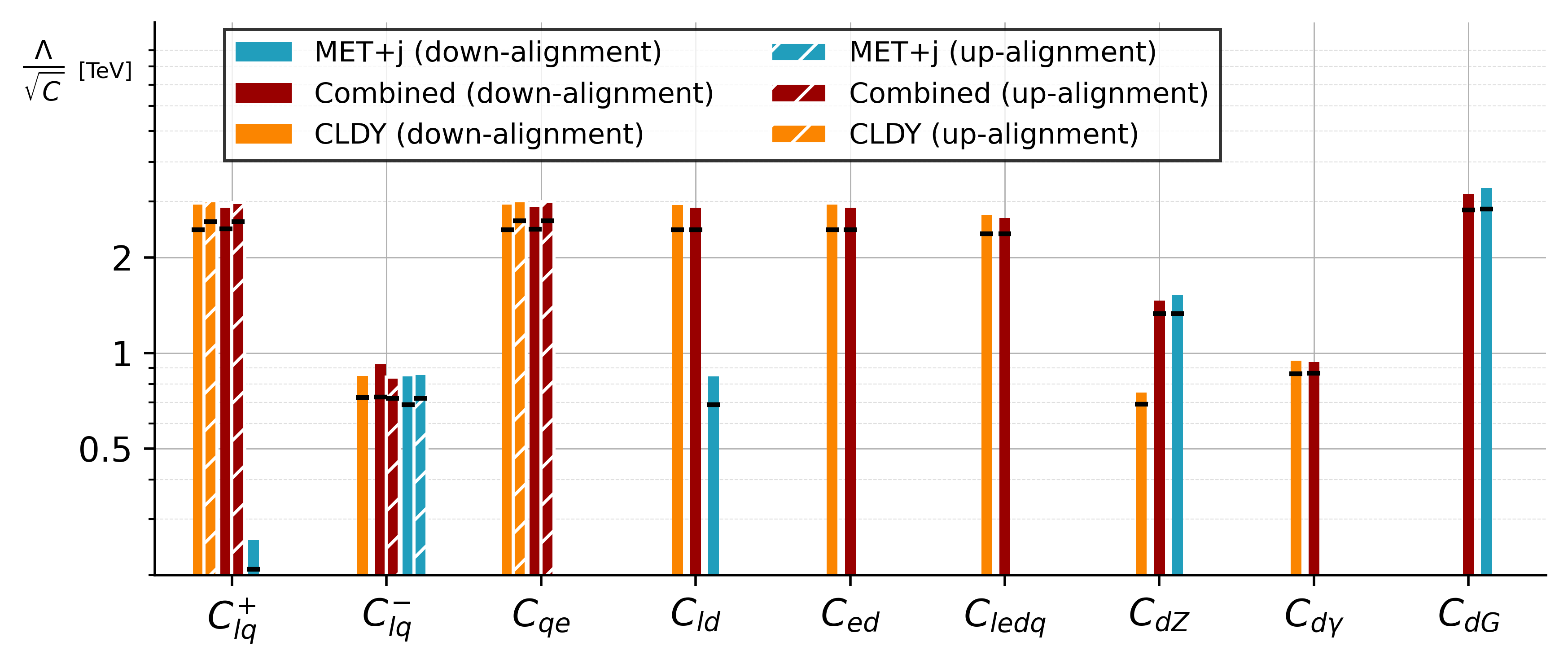}
  \caption{95\% limits on $\Lambda/\sqrt{C}$ in the lepton-flavor universal scenario for {$i,j=2,3$}, see Fig.~\ref{fig:results_LU_12}.}
  \label{fig:results_LU_23}
\end{figure}

The Drell-Yan data used in this analysis given in Table \ref{tab:data_sets} probes new physics scales in excess of 10 TeV, for operators involving the
first and second generation quarks. As parton luminosities for $b$-quarks are smaller, corresponding operators are constrained less accordingly, see Figs.~\ref{fig:results_LU_13}
and \ref{fig:results_LU_23}.
Limits for different lepton flavor assumptions are comparable, cf. Fig \ref{fig:results_LFV_12}.

Constraints from rare decays of kaons, charm and beauty are generically stronger for semileptonic four-fermion operators, with the notable
exception of right-handed taus, as there is no phase-space for the first two generation quark FCNCs \footnote{The LFV decay $D^0 \to e \tau$ is the sole exception to this, but there is no data available~\cite{ParticleDataGroup:2024cfk}.}
and they are not covered by an $SU(2)_L$-link with dineutrino data.
In addition, we find that flavor constraints are beneficial for the dipole operators, for which limits improve in the combination by up to a factor of three, see Table~\ref{tab:combined_constraints}.

The presence of light degrees of freedom, such as sterile neutrinos, would affect the missing energy observables only. Hence, breakdown of correlations between charged and neutral
Drell Yan indicates that new physics is not captured by SMEFT, but rather requires to take into account further degrees of freedom. We therefore encourage further MET data taking and to include both charged and
neutral dilepton production in global analyses.

An estimation of the reach in new physics scales at future hadron colliders is shown in Table~\ref{tab:FutureNP}. Compared to existing LHC data, the reach improves at the HL-LHC by $\sim 1.5$, the HE-LHC
 by $\sim 3$ and at the FCC-hh by $\sim 8$. Our study shows also that the largest tail is not always the one with the best sensitivity, and the position of the highest bin needs to be optimized, see e.g. 
Figs.~ \ref{fig:RatioPlot} and \ref{fig:NPreach_uc_bs}. One also observes the flavor dependence of the optimal bin. Ideally, the binning should be tuned to the
flavor content of an operators, to maximize the reach.

We focused in this work on quark FCNC processes, however our study is generalizable to flavor diagonal transitions, with $i=j$. The main difference are the quark PDFs involved and
the presence of interference terms with the SM for some four-fermion operators, see Sec.~\ref{sec:cross_sections}. In this (flavor-diagonal) case also contributions from dimension-8 operators become potentially relevant.
In addition, one could also perform a global fit assuming flavor textures for the quarks, which allows for correlations between the different generations. Corresponding extensions of our analysis are of interest but beyond the scope of this work.

As more data is harvested from the LHC during Run 3 and the high luminosity phase, making use of correlations between different and complementary data-sets, as well as combinations between collider and rare decay data is a rewarding avenue for testing the Standard Model deeper. We look forward to future data from flavor, high-$p_T$ and beyond.

\begingroup
\renewcommand{\addcontentsline}[3]{}
\begin{acknowledgments}
 We thank Joachim Brod, Emmanuel Stamou, and Dominik Suelmann for useful discussions. LN is supported by the doctoral scholarship program of the {\it Studienstiftung des Deutschen Volkes}. This research was supported in part by grant NSF PHY-2309135 to the Kavli Institute for Theoretical Physics (KITP).
\end{acknowledgments}
\endgroup

\renewcommand{\thesubsection}{\Alph{subsection}}

\numberwithin{equation}{subsection} 
\renewcommand{\theequation}{\thesubsection.\arabic{equation}}

\section{Appendix}
\label{sec:Appendix}

\subsection{Partonic cross sections}
\label{app:partonic_xsec}

In this section, we outline the parametrization of the partonic cross section, based on Refs.~\cite{Hiller:2024vtr},\cite{Allwicher:2022gkm} for the DY process with dineutrinos and charged leptons, respectively.

The partonic cross sections $\hat \sigma(q\bar q \rightarrow \ell^+ \ell^-)$ in Eq.~\eqref{eqn:xsec_ll} in the high energy limit $\hat s \gg M_z^2$ read 
\begin{align}
  \hat \sigma_{\text{SM}}\left(q\bar q \rightarrow \ell \ell \right) &= \frac{4 \pi \alpha^2}{9 \hat s}\left(Q_q^2 + \frac{\left( {\epsilon_L^{q}}^2 + {\epsilon_R^{q}}^2 \right) \left( {\epsilon_L^{\ell }}^2 + {\epsilon_R^{\ell}}^2 \right)}{4 c_W^4 s_W^4 } -\frac{ Q_q\left(\epsilon_L^{q} + \epsilon_R^{q} \right) \left(\epsilon_L^{\ell} + \epsilon_R^{\ell} \right) }{ 2 c_W^2 s_W^2} \right) \,, \label{eqn:DY_SM_xsec} \\
  \hat \sigma_{4F,\text{lin}}\left(q\bar q \rightarrow \ell \ell\right) &= \frac{ \alpha }{216 c_W^2 \Lambda^2} \label{eqn:DY_4F_xsec_lin} \,, \\ 
  \hat \sigma_{4F,\text{quad}}\left(q\bar q \rightarrow \ell \ell\right) &= \frac{ \hat s}{ 144 \pi \Lambda^4} \label{eqn:DY_4F_xsec} \,, \\ 
  \hat \sigma_{\text{EW}}\left(q\bar q \rightarrow \ell \ell\right) &= \frac{1 - 4 s_W^2 + 8 s_W^4}{4c_W^2 s_W^2}\frac{v^2 \alpha }{18 \Lambda^4} \,, 
  \label{eqn:DY_EW_xsec} 
\end{align}
where $\alpha = e^2/4\pi$ is the electromagnetic fine structure constant, $v \sim 246\, \mathrm{GeV} $, $\Lambda$ the NP scale and $\hat s$ the partonic center of mass energy. The $Z$-couplings $\epsilon$ are given by 
\begin{equation}
  \epsilon_X^{f} = T^3_{X} - s_W^2 Q_f \,,
  \label{eqn:Z_couplings_SM}
\end{equation}
in the SM, for a fermion $f = q, \ell$ with charge $Q_f$, weak isospin $T^3_{X}$ and chirality $X = L/R$.
Typically, experiments probe the $q^2$-spectrum, which is related to $\hat{s}$ by the equation ${q^2 = \hat{s}}$ at LO. For CLDY at LO, gluons do not contribute as initial states.

The differential cross-sections $\frac{\mathrm{d}\hat \sigma( q_i g \rightarrow \nu \bar \nu q_j)}{\mathrm{d} p_T}$ of the dineutrino process in Eq.~\eqref{eqn:xsec_nunu} read
\begin{align} 
  \frac{\mathrm{d}\hat \sigma_{\text{SM}}( q g \rightarrow \nu \bar \nu q)}{\mathrm{d} P_T } &= \frac{ \alpha_s {\cal{B}}( Z \rightarrow \nu \bar \nu) M_Z^2\left( {\epsilon_L^{q}}^2 + {\epsilon_R^{q}}^2 \right)}{6 v^2 }\frac{1}{\hat s^{3/2} } \frac{ x^2 + 4 }{ x \sqrt{ 1- x^2} } \,, \label{eqn:parton_SM} \\
  \frac{\mathrm{d}\hat \sigma_{4F}( q g \rightarrow \nu \bar \nu q)}{\mathrm{d} P_T } &= \frac{5\alpha_s }{432 \sqrt{2} \pi^2 } \frac{1}{\Lambda^4}\sqrt{\hat s} (1-x)^{3/2} \,, \label{eqn:parton_4F} \\ 
  \frac{\mathrm{d} \hat \sigma_{\text{EW}}( q g \rightarrow \nu \bar \nu q)}{\mathrm{d} P_T } &= \frac{ \alpha_s {\cal{B}}\left( Z \rightarrow \nu \bar \nu\right) }{6 } \frac{v^2}{\Lambda^4} \frac{1}{\sqrt{\hat s} } \frac{ x}{\sqrt{ 1- x^2}} \,, \label{eqn:parton-EW} \\ 
  \frac{\mathrm{d} \hat \sigma_{G}( q g \rightarrow \nu \bar \nu q)}{\mathrm{d} P_T } &= \frac{{\cal{B}}\left( Z \rightarrow \nu \bar \nu\right) }{96 \pi} \frac{1}{\Lambda^4} \sqrt{\hat s} \frac{x}{\sqrt{ 1- x^2} } \,, \label{eqn:parton-G}
  \end{align}
where~\eqref{eqn:parton_SM},~\eqref{eqn:parton-EW} and~\eqref{eqn:parton-G} are given in the narrow width approximation (NWA), with ${q^2 = M_Z^2}$. They are expanded for $ M_Z^2\ll\hat s ,4 P_T^2$, while Eq.~\eqref{eqn:parton_4F} is expanded around $x\sim 1$.
Furthermore, $x = 2 p_T \, / \sqrt{\hat s}$, $\alpha_s$ is the strong coupling constant, $M_Z$ the $Z$-boson mass and ${\cal{B}}( Z \rightarrow \nu \bar \nu)$ the branching ratio of the $Z$-boson to invisible final states\footnote{The definitions of $\epsilon_X^{f}$ are related by a factor of $ -2 M_Z /v$ to the definitions used in Ref.~\cite{Hiller:2024vtr}.}. 
For the four-fermion interference term and the non-expanded version of the formulas, see Ref.~\cite{Hiller:2024vtr}.
The variable $p_T$ denotes the transverse momentum of the dineutrino pair which corresponds to $\emiss$ at LO, while $q^2 \neq \hat s$ is the corresponding invariant mass.

The total cross section receives additional contributions from $q\bar q$-channels, which are related to the $q\bar g$ channels by crossing symmetry. More details can be found in Ref.~\cite{Hiller:2024vtr}.

The full energy enhancement of $\mathrm{d} \hat \sigma_{G}$ breaks the naive energy scaling, which can be traced back to the longitudinal modes of the $Z$-boson.
This is further explained in App.~\ref{app:G_dipoles}.

In the experiment, the MET+j process is measured differentially as an $\emiss$-spectrum, where $\emiss$ is the a sum of the transverse momenta off all visible final states, including the leading jet $p_T$ as well as additional softer jets. 
The total hadronic cross section can then be written as 
\begin{equation}
  \label{eqn:Hadronic_xsec}
  \mathrm{d}\sigma = \sum_{i,j}\int \frac{\mathrm{d}\tau }{\tau} \mathrm{d} \hat \sigma_{i j}(\tau s,...) L_{ij}(\tau,\mu_F^2) \,,
\end{equation}
where $i,j = u,d,s,c,b,g$ and the parton luminosity functions are defined as 
\begin{equation}
  L_{i j}(\tau,\mu_F^2) = \tau \int_{\tau}^{1} \frac{\mathrm{d}x}{x} \left[ f_i(x,\mu_F^2) f_{\bar j}(\tau/x,\mu_F^2) + f_j(x,\mu_F^2) f_{\bar i}(\tau/x,\mu_F^2) \right] \,,
  \label{eqn:PLFs}
  \end{equation} 
with the proton PDFs $f_i(x,\mu_F^2)$ and the factorization scale $\mu_F$.

\subsection{Energy enhancement of the gluon dipole operators}
\label{app:G_dipoles}

To examine the energy enhancement of the gluon dipole operators we work in the NWA and only consider on-shell $Z$-bosons.
The energy enhancement of $C_{\underset{ij}{uG}}$ and $C_{\underset{ij}{dG}}$ observed in~\cite{Hiller:2024vtr} for MET+j can be understood by considering the longitudinal modes of the $Z$-boson. The fraction of this polarization grows with increasing momentum and dominates in the high energy regime.

This can be explicitly shown by using the Goldstone equivalence theorem~\cite{PhysRevD.10.1145,PhysRevD.16.1519}, which states that amplitudes for longitudinal polarized vector bosons are equivalent to their respective Goldstone modes $\pi$ in the high energy limit.

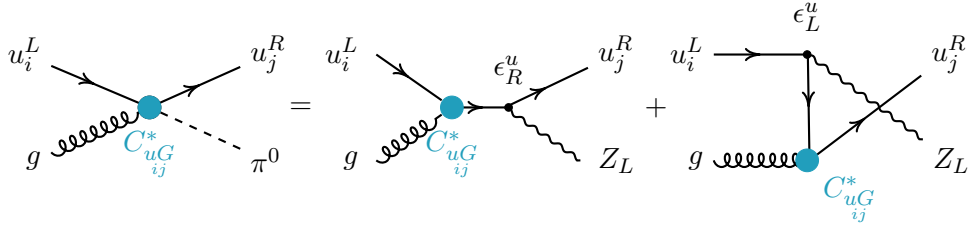
\begin{figure}[htb]
  \xdefinecolor{dRed}{RGB}{153, 0, 0}
\xdefinecolor{dOrange}{RGB}{251, 133, 0}
\xdefinecolor{dBlue}{RGB}{33, 158, 188}
\xdefinecolor{dGrey}{RGB}{100, 100, 100}

\begin{tikzpicture}[scale=1] \begin{feynman}
  
    \vertex [crossed dot](ww1){};
    \vertex [left=1.cm of ww1] (ml);
    \vertex [right=1.cm of ww1] (mr);
    \vertex [above=0.7cm of ml] (u1);
    \vertex [left=0.3cm of u1] (u) {\(u^L_i\)};
    \vertex [below=0.7cm of ml] (u2);
    \vertex [left=0.3cm of u2] (u3) {\(g\)};
    \vertex [above=0.7cm of mr] (l1);
    \vertex [right=0.2cm of l1] (l) {\( u^R_j\)};
    \vertex [below=0.7cm of mr] (l2);
    \vertex [right=0.2cm of l2] (l3) {\(\pi^0\)};

    \vertex [right=4.0cm of ww1] (ww3){};
    \vertex [right=0.7cm of ww3] (ww4);
    \vertex[left=1cm of ww3] (ml2);
    \vertex[above=0.7cm of ml2] (gg1);
    \vertex[left=0.1cm of gg1] (g1) {\(u^L_i\)};
    \vertex[below=0.7cm of ml2] (qq1);
    \vertex[left=0.1cm of qq1] (q1) {\(g\)};
    \vertex[right=1cm of ww4] (mr2);
    \vertex[above=0.7cm of mr2] (qq2);
    \vertex[right=0.1cm of qq2] (q2) {\(u^R_j\)};
   \vertex[below=0.7cm of mr2] (h1);
   \vertex[right=0.2cm of mr2] (h2);
   \vertex[right=0.1cm of h1] (nubar) {$Z_L$};

   \vertex [right=4.0cm of ww4] (ww5);
   \vertex [right=0.7cm of ww5] (ww6);
   \vertex[left=1cm of ww5] (ml3){};
   \vertex[above=0.7cm of ml3] (gg2);
   \vertex[left=0.1cm of gg2] (g2) {\(u^L_i\)};
   \vertex[below=0.7cm of ml3] (qq2);
   \vertex[left=0.1cm of qq2] (q3) {\(g\)};

  \vertex[right=1.6cm of g2] (ww6);
  \vertex[right=1.5cm of q3] (ww7);
  \vertex[right= 1.5cm of ww6](g3){\(u^R_j\)};
  \vertex[right= 1.5cm of ww7](ZL){\(Z_L\)};

    \diagram* {
    (u) -- [thick,fermion] (ww1),
    (u3) -- [thick,gluon] (ww1),
    (l3) -- [thick,scalar] (ww1),
    (ww1) -- [thick,fermion ] (l),

   (ww3) -- [thick,fermion] (ww4) -- [thick, fermion] (q2),
   (gg1) -- [thick,fermion] (ww3),
   (qq1) -- [thick, gluon] (ww3),
    (ww4) -- [thick,boson] (h1),

    (g2) -- [thick,fermion] (ww6) -- [thick, boson] (ZL),
    (q3) -- [thick,gluon] (ww7),
    (ww7) -- [thick, fermion] (g3),
    (ww6) -- [thick,fermion] (ww7),

      };
  


  
  
    \draw ( 2,0)node{$= $};
    \draw ( 6.7,0)node{$+$};
    \draw[fill = dBlue, dBlue ] (0,-0.0) circle (0.15) node[below = 5pt] {$C_{\underset{ij}{uG}}^{*} $};
    \draw[fill=dBlue, dBlue] (4,-0.0) circle (0.15) node[below = 5pt] {$C_{\underset{ij}{uG}}^{*} $};
    \draw[fill=dBlue, dBlue] (8.7,-0.7) circle (0.15) node[below right = 3pt] {$C_{\underset{ij}{uG}}^{*} $};

    \draw[fill = black, black ] (4.75,-0.0) circle (0.05) node[above = 5pt] {$\epsilon^u_R $};
    \draw[fill = black, black ] (8.7,0.7) circle (0.05) node[above = 5pt] {$\epsilon^u_L $};
  %

  \end{feynman} \end{tikzpicture}
  
  \caption{Goldstone equivalence theorem for $ u^L_i g \to u^R_j Z_L$, where $ R,L$ denotes the chirality of the quarks and $Z_L$ denotes the longitudinal mode of the $Z$-boson . 
  On the left hand side we show the contribution to $u^L_i g \to u^R_j \pi^0$, where $\pi^0$ is the neutral Goldstone boson. 
  On the right hand side, the diagrams contributing to $u^L_i g \to u^R_j Z_L$ are shown, where the $\epsilon^u_X$ are the $Z$-boson SM couplings defined in Eq.~\eqref{eqn:Z_couplings_SM}. 
  Additional diagrams contribute on the left hand side, which are however suppressed by the Yukawa couplings and thus neglected here.}
  \label{fig:GET}
\end{figure}

As an example, we consider the process $u^L_i g \to u^R_j Z$, where $R,L$ denotes the chirality of the quark, with the operator $O_{uG}$.
The Goldstone equivalence theorem for this process is schematically illustrated in Fig.~\ref{fig:GET}.

Expanding the Higgs doublet around the vacuum expectation value $v$ as
\begin{equation}
\varphi = \begin{pmatrix} \pi^+ \\
 \frac{v+ h + i \pi^0}{\sqrt{2}} 
\end{pmatrix},
\end{equation}
allows us to expand the operator $O_{uG}$ as
\begin{equation}
  O_{\underset{ij}{uG}} = \bigl(\bar u_i \sigma^{\mu\nu} T^A P_R u_j \bigr) \frac{(v+ h -i \pi^0)}{\sqrt{2}} G_{\mu\nu}^A - \bigl(\bar d_i \sigma^{\mu\nu} T^A P_R u_j \bigr) \pi^- G_{\mu\nu}^A \,,
  \label{eqn:CuG_expanded}
\end{equation}
where $P_{R,L} = (1 \pm \gamma^5) \; /2$ are the right- and left-handed projection operators, respectively, and $\pi^0$ ($\pi^\pm$) denote the neutral (charged) $SU(2)$ Goldstone bosons. 
Employing naive dimensional analysis, one would expect that the term proportional to $v$ contributing to the $gq \bar q$-Vertex is not fully energy enhanced. \\
However, an alternative approach to this calculation is given by the Goldstone equivalence theorem, i.e. the left hand side of Fig.~\ref{fig:GET}.
There, the contact term generated for $ u^L_i g \to u^R_j \pi^0$ in Eq.~\eqref{eqn:CuG_expanded} is fully energy enhanced in naive dimensional analysis and it involves no factors of $v$.
The $Z$-boson couplings $\epsilon^u_X$ are inherently absent in this calculation, but the universal combination $\left(\epsilon^u_L -\epsilon^u_R\right)^2$ appears implicitly as a factor in the calculation.
The full calculation for $u^L_i g \to u^R_j Z_L$ is more involved, as the diagrams on the right hand side of Fig.~\ref{fig:GET} involve the SM $Z$-boson vertex $\epsilon^u_X$, see Eq.~\eqref{eqn:Z_couplings_SM}, and a $s$-and a $t$-channel.
As can be seen in the right hand side of Fig.~\ref{fig:GET} the insertion of $\mathcal{O}_{uG}$ leads to a chirality flip, which means the different channels are proportional to either $\epsilon^u_L$ or $\epsilon^u_R$. 
In the high energy limit, the longitudinal polarization vectors are given by $\epsilon^{\mu}(q^{\mu}) = \frac{q^{\mu} }{M_Z}$, where $q^{\mu} $ is the $Z$-boson four momentum.
Inserting this leads to  cancellations and a contact term proportional to $\left( {\epsilon^u_L -\epsilon^u_R}\right)^2 \propto \left(T^3\right)^2 $ survives.
The full calculation is given in Ref.~\cite{Hiller:2024vtr}, where in Eq.~(B64) this term also apperas and  furthermore scales with $\hat s$.
An analogous discussion holds for processes induced by $C_{\underset{ji}{uG}}$ and down-type quark processes induced by $C_{{\underset{ij}{dG}}},C_{{\underset{ji}{dG}}} $.
Similarly for $O_{uB}$ and $O_{uW}$ this reads 
\begin{align}
 & O_{\underset{ij}{uB}} = \bigl(\bar u_i \sigma^{\mu\nu} P_R u_j \bigr) \frac{(v+ h -i \pi^0)}{\sqrt{2}} B_{\mu\nu} - \bigl(\bar d_i \sigma^{\mu\nu} P_R u_j \bigr) \pi^- B_{\mu\nu}  \,,
 \label{eqn:CuB_expanded} \\
 & O_{\underset{ij}{uW}} = \bigl(\bar u_i \sigma^{\mu\nu} P_R \tau^I u_j \bigr) \frac{(v+ h -i \pi^0)}{\sqrt{2}} W^I_{\mu\nu} - \bigl(\bar d_i \sigma^{\mu\nu} P_R u_j \bigr) \pi^- W^I_{\mu\nu}\,. 
 \label{eqn:CuW_expanded}
\end{align}
Both would contribute fully energy enhanced to $pp \to Z \pi^0$, which is connected through the Goldstone equivalence theorem to $p p \to Z Z_L$.
However, we only consider ${ p p \to Z g}\; {( Z \to \nu \bar \nu)}$, which does not benefit from this and therefore the EW dipoles scale proportional to $v$ in our analysis.

\subsection{Auxiliary limits and plots}
\label{app:auxiliary_limits}

In this appendix we provide auxiliary information on the fits and the study of future reaches.

\begin{table}
  \setlength{\tabcolsep}{15pt}
  \centering
  \begin{tabular}{l |l l l}
    & $i,j=1,2$ & $i,j=1,3$ & $i,j=2,3$ \\
    \toprule 
    $C_{\underset{ij}{uB}}$ & $[-0.34, 0.34]$ &  &  \\
    $C_{\underset{ij}{uW}}$ & $[-0.21, 0.21]$ &  &  \\
    $C_{\underset{ij}{uG}}$ & $[-0.020, 0.020]$ &  &  \\
    $C_{\underset{ij}{dB}}$ & $[-0.34, 0.34]$ & $[-0.76, 0.76]$ & $[-1.0, 1.0]$ \\
    $C_{\underset{ij}{dW}}$ & $[-0.24, 0.24]$ & $[-0.46, 0.46]$ &  $[-0.68, 0.68]$ \\
    $C_{\underset{ij}{dG}}$ & $[-0.032, 0.032]$ & $[-0.041, 0.041]$ & $[-0.10, 0.10]$ \\
    $C_{\underset{ij}{lq}}^{(1)}$ (up) & $[-0.023, 0.0095]$ & $[-0.35, 0.35]$ & $[-0.72, 0.70]$ \\
    $C_{\underset{ij}{lq}}^{(1)}$ (down) & $[-0.012, 0.020]$ & $[-0.17, 0.33]$ & $[-0.38, 0.80]$ \\
    $C_{\underset{ij}{lq}}^{(3)}$ (up) & $[-0.023, 0.0094]$ & $[-0.35, 0.35]$ & $[-0.72, 0.70]$ \\
    $C_{\underset{ij}{lq}}^{(3)}$ (down) & $[-0.02, 0.012]$ & $[-0.33, 0.16]$ & $[-0.80, 0.38]$ \\
    $C_{\underset{ij}{qe}}$ (up)& $[-0.015, 0.018]$ & $[-0.078, 0.078]$ & $[-0.11, 0.12]$ \\
    $C_{\underset{ij}{qe}}$ (down) & $[-0.011, 0.013]$ & $[-0.076, 0.076]$ & $[-0.12, 0.12]$ \\
    $C_{\underset{ij}{lu}}$ & $[-0.023, 0.023]$ &  &  \\
    $C_{\underset{ij}{ld}}$ & $[-0.029, 0.029]$ & $[-0.076, 0.076]$ & $[-0.12, 0.12]$ \\
    $C_{\underset{ij}{eu}}$ & $[-0.023, 0.023]$ &  &  \\
    $C_{\underset{ij}{ed}}$ & $[-0.029, 0.029]$ & $[-0.076, 0.076]$ & $[-0.12, 0.12]$ \\
    $C_{\underset{ij}{ledq}}$ & $[-0.034, 0.034]$ & $[-0.089, 0.088]$ & $[-0.14, 0.14]$ \\
    $C_{\underset{ij}{lequ}}^{(1)}$ & $[-0.027, 0.027]$ &  &  \\
    $C_{\underset{ij}{lequ}}^{(3)}$ & $[-0.012, 0.012]$ &  &  \\
    \hline
  \end{tabular}
  \caption{95\% credible intervals for the SMEFT coefficients in the lepton-flavor universal scenario assuming $\Lambda= 1$ TeV.
  For coefficients with quark-doublet currents $C_{lq}^{(1)}, C_{lq}^{(3)}, C_{qe}$ ranges depend on the gauge-mass-alignment which is specified in parentheses.}
  \label{tab:CL_LU}
\end{table}

\begin{figure}[h]
  \centering
  \includegraphics[width=0.48\textwidth]{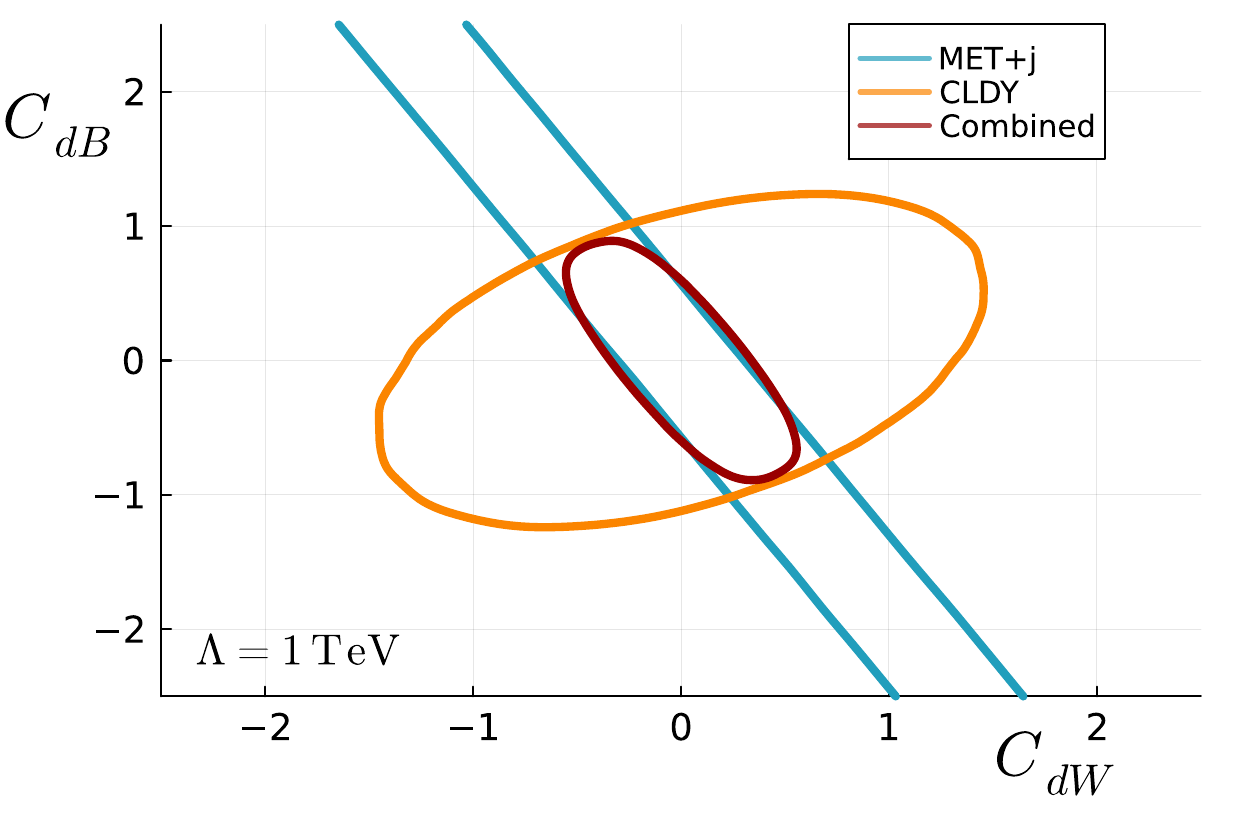}
  \includegraphics[width=0.48\textwidth]{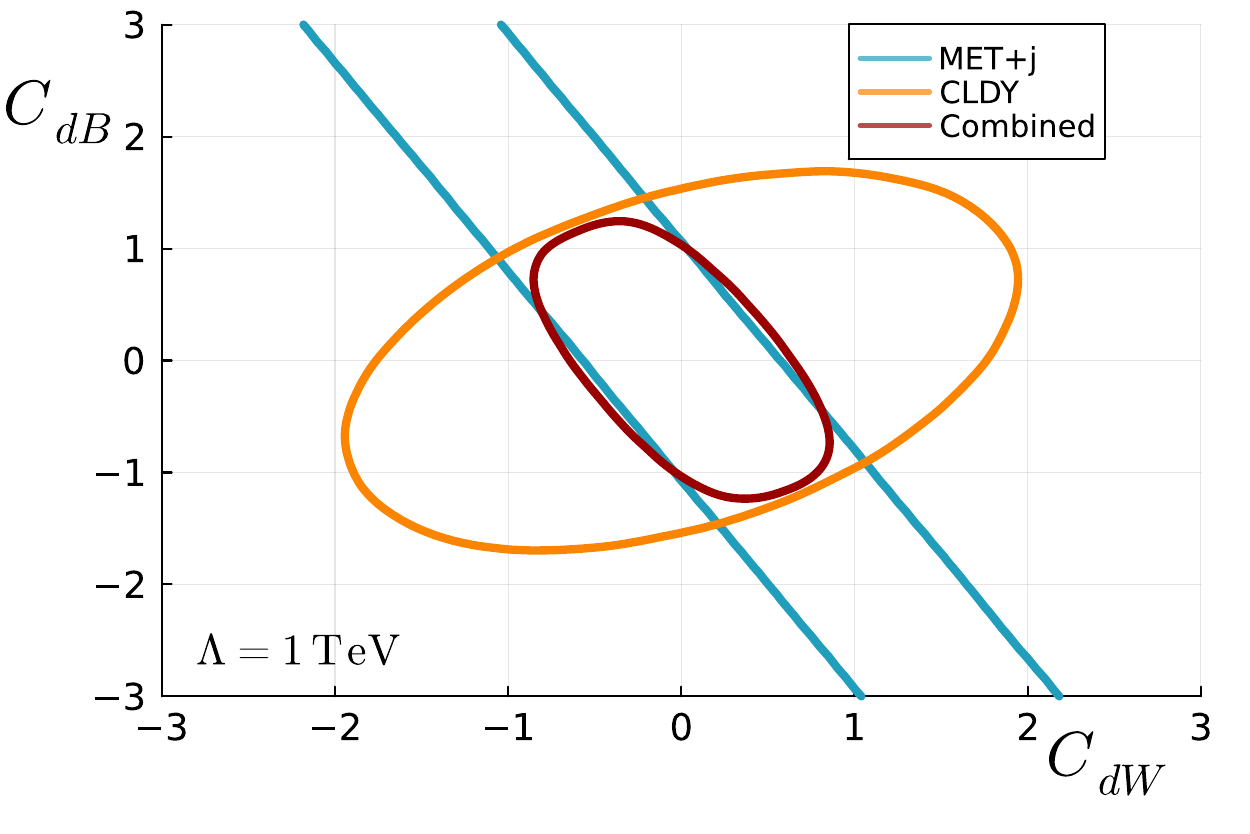}
  \caption{95\% credible contours for the dipole coefficients $C_{dW}$ and $C_{dB}$ for the quark indices $i,j=1,3$ (left) and for $i,j=2,3$ (right) for $\Lambda= 1$ TeV, see Fig.~\ref{fig:results_dipoles_12}.}
  \label{fig:results_dipoles_app}
\end{figure}

\begin{figure}[h]
  \centering
  \includegraphics[width=0.8\textwidth]{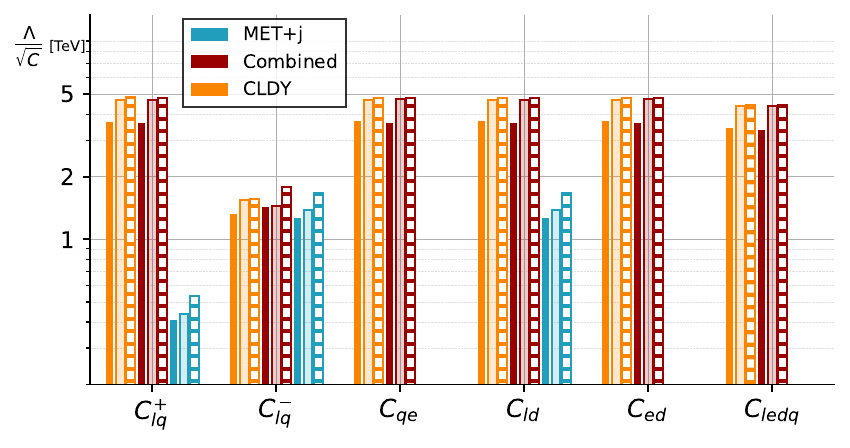}
  \caption{95\% limits on $\Lambda/\sqrt{C}$ in the lepton-flavor universal (solid), lepton-flavor violating (shaded) and democratic (striped) scenario for the quark indices $i,j=1,3$ from the combined fit (red) 
  and the individual fits of the CLDY (orange) and MET+j (blue) in  down-alignment, 
  see Fig.~\ref{fig:results_LFV_12}.}
  \label{fig:results_LFV_13}
\end{figure}

\begin{figure}[h]
  \centering
  \includegraphics[width=0.8\textwidth]{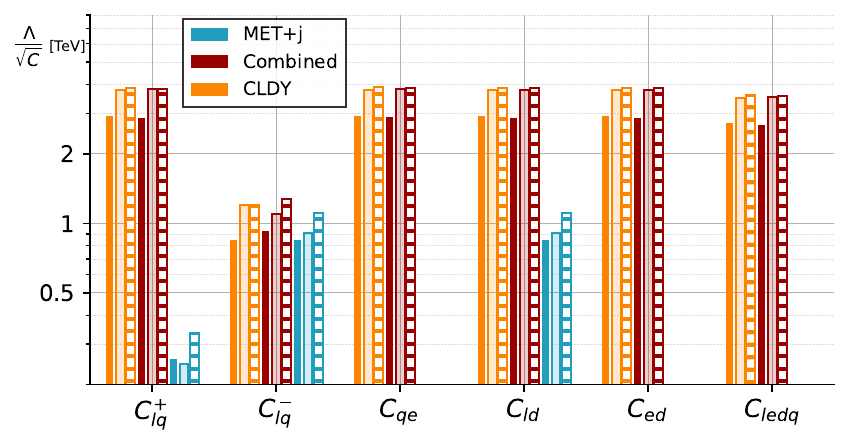}
  \caption{95\% limits on $\Lambda/\sqrt{C}$ in the lepton-flavor universal (solid), lepton-flavor violating scenario (shaded) and democratic (striped) scenario  for the quark indices $i,j=2,3$ from the combined fit (red) 
  and the individual fits of the CLDY (orange) and MET+j (blue) in down-alignment, 
  see Fig.~\ref{fig:results_LFV_12}.}
    \label{fig:results_LFV_23}
\end{figure}

\clearpage

\bibliography{references}

\end{document}